\documentclass[aps,pre,floatfix,superscriptaddress,twocolumn,showkeys,10pt]{revtex4-2}
\usepackage{color}
\usepackage{graphicx}
\usepackage{amsmath, amsthm, amssymb}
\usepackage[colorlinks,citecolor=red,urlcolor=blue]{hyperref}
\newcommand{\be}{\begin{equation}}
\newcommand{\ee}{\end{equation}}
\newcommand{\bea}{\begin{eqnarray}}
\newcommand{\eea}{\end{eqnarray}}

\begin{document}
\title{The phase transition from  nematic to  high-density disordered phase in a system  of hard rods on a lattice}

\author{Aagam Shah}
\email{aagam.shah@students.iiserpune.ac.in}
\affiliation{Indian Institute of Science Education and Research Pune, Dr. Homi Bhabha Road, Pashan,  Pune 411008, India}
\author{Deepak Dhar}
\email{deepak@iiserpune.ac.in}
\affiliation{Indian Institute of Science Education and Research Pune, Dr. Homi Bhabha Road, Pashan,  Pune 411008, India}
\author{R. Rajesh} 
\email{rrajesh@imsc.res.in}
\affiliation{The Institute of Mathematical Sciences, C.I.T. Campus, Taramani, Chennai 600113, India}
\affiliation{Homi Bhabha National Institute, Training School Complex, Anushakti Nagar, Mumbai 400094, India}

\date{\today}

\begin{abstract}
A system of hard rigid rods of length $k$ on hypercubic lattices is known to undergo two phase transitions when chemical potential is increased: from a low density isotropic phase to an intermediate density nematic phase, and  on further increase  to a high-density phase with no orientational order. In this paper, we argue that, for large $k$, the second phase transition is a first order transition with a discontinuity in density in all dimensions greater than $1$. We show that the chemical potential at the transition is  $\approx  k \ln [k /\ln k]$ for large $k$,   and that the density of uncovered sites drops from a value  $ \approx (\ln k)/k^2$ to a value of order $\exp(-ak)$, where $a$ is some  constant, across the transition. We conjecture that these results are  asymptotically exact, in all dimensions $d\geq 2$. We also present evidence of coexistence of  nematic and disordered phases  from Monte Carlo simulations  for rods of length $9$ on the square lattice.
\end{abstract}

\maketitle

\section{\label{sec:intro}Introduction}

The study of entropy driven phase transitions in systems of long hard rods is one of the classic problems of Statistical Mechanics. It has a long history, starting with Onsager establishing an isotropic-nematic phase transition in a solution of long thin rods in three dimensions~\cite{1949-o-nyas-effects}, and Zwanzig developing a virial expansion  for rods on lattices~\cite{1963-z-jcp-first}. 
The model of hard rods  are good minimal models for many  phase transitions, e.g.,  those observed in aqueous solutions tobacco mosaic viruses~\cite{1989-fmcm-prl-isotropic}, liquid crystals~\cite{1995-oup-gp-physics},  carbon nanotube nematic gels~\cite{2004-iadzly-prl-nematic}, etc.

\begin{figure}
\includegraphics[width=\columnwidth]{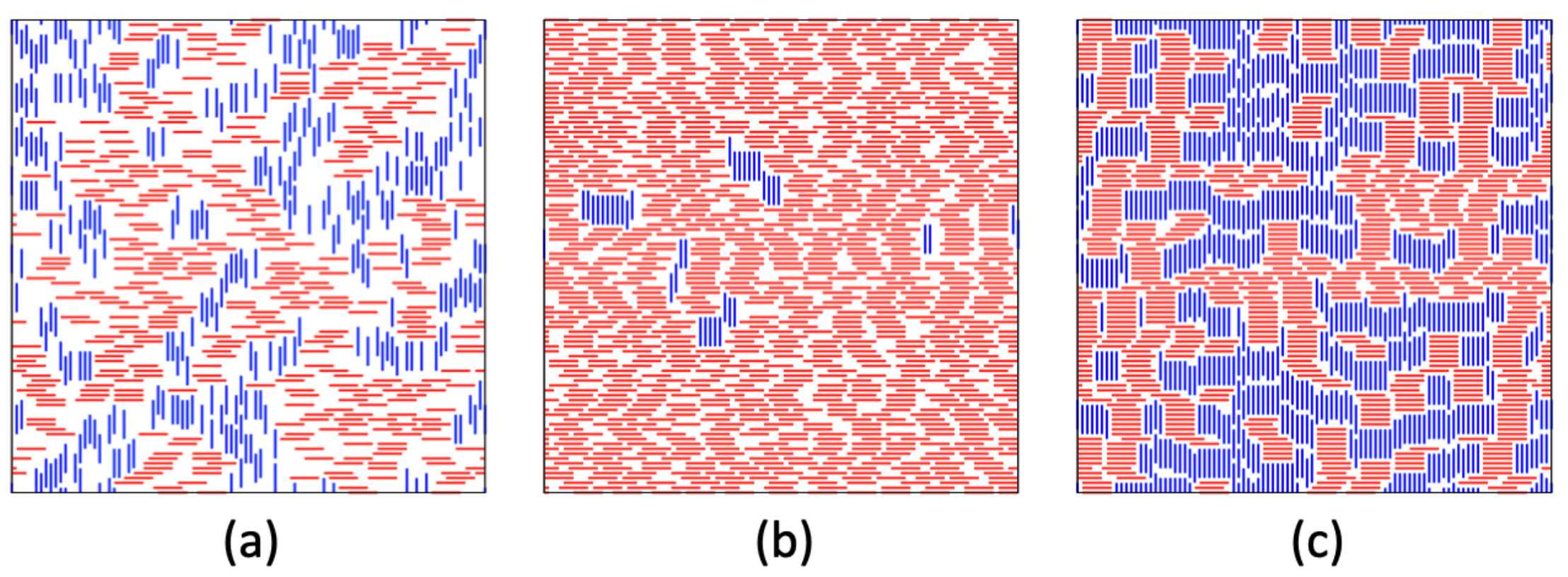}
\caption{\label{fig:schematicsnapshots}(Color online) Typical snapshots of the three phases at different densities for rods of length $7$.  (a) The low density disordered phase, (b) the intermediate density nematic phase, and (c) the HDD phase. $x$-mers are shown in red and $y$-mers in blue.}
\end{figure}

In this paper, we focus on lattice models for mono-dispersed straight rigid rods. On  a $d$-dimensional hyper-cubic lattice, rods can  orient only in one of the $d$ directions. A $k$-mer will refer to a rod of length $k$ that occupies $k$ consecutive lattice sites along any one of the lattice directions. Two rods cannot overlap.  With increasing density, it is known that, for large enough $k$, the system of $k$-mers undergoes transitions from a low density orientationally disordered phase to an intermediate density nematically order phase to a high density disordered (HDD) phase where the nematic order is lost (see Fig.~\ref{fig:schematicsnapshots} for an illustration of these phases)~\cite{2007-gd-epl-on}. The first transition from the disordered to nematic phase is expected to lie in the Ising~\cite{2008-mlr-epl-determination,2008-mlr-jcp-critical,2009-fv-epl-restricted} or more generally Potts universality class~\cite{2008-mlr-epl-determination,2008-mlr-jcp-critical,2008-mlr-pa-critical}, depending on the number of  different possible directions of nematic order. The transition has been rigorously established to exist in two dimensions~\cite{2013-dg-cmp-nematic}, and  is also seen in the exactly soluble case of $k$-mers on tree-like lattices~\cite{2011-drs-pre-hard}. It has also been shown that  machine learning can be  used to detect this phase transition~\cite{padavala2021machine}.

The second transition from the nematic to the HDD phase is much  less studied. In Monte Carlo simulations, even to establish the existence of the second transition is nontrivial, as the approach to  thermal equilibrium at high densities becomes very slow.  In the usual algorithms using local evaporation and deposition moves, the  states at high densities are sampled inefficiently  due to the presence of highly jammed long-lived metastable states. This difficulty is reduced substantially by a recently introduced strip update cluster algorithm  based on simultaneously updating all the sites in a strip based on transfer matrix calculations~\cite{2012-krds-aipcp-monte,2013-krds-pre-nematic}.

The nature of the second transition as well as the nature of the HDD phase is not settled yet.  There is some indication of the HDD phase having  power law orientation-orientation correlations~\cite{2013-krds-pre-nematic}. The results of Monte Carlo simulations of systems upto size $952 \times 952$ for $k=7$, were consistent with a continuous transition in a non-Ising universality class~\cite{2013-krds-pre-nematic}.  However, an exact solution 
of soft repulsive rods on a tree-like lattice~\cite{2013-kr-pre-reentrant} suggests a continuous transition but in the Ising universality class.

This transition has been studied more recently by Vogel et. al. by using  an interesting new measure to study the Monte Carlo data~\cite{vogel2017phase,vogel2020alternative}. The size of the file storing the time series of configurations is reduced in size using a  zipping program.  The ratio of the reduced file size to original size, termed as mutability, changes with chemical potential. It is argued that the maxima and minima of mutability can be used as markers of phase transitions in the system being modeled.  But there  is no simple relationship between mutability and standard thermodynamic variables. The precise value would also depend on the details of the zipping program used.  If the system undergoes slow relaxation, then nearby configurations are more similar, and mutability will be low.  This suggests that mutability tracks the inverse relaxation time, but near the isotropic-nematic transition  mutability  actually shows a maximum.   Also, if there is a first order transition, we would expect the mutability to show a minimum in the middle of the two-phase-coexistence region, if the boundary between the region fluctuates.   The transitions, for rods upto size $k=10$, were found to be consistent with a continuous transition.

In another study~\cite{chatelain2020absence}, corner transfer matrix  renormalization group technique was used to study the phase transitions  in hard rods.  While this technique gives rather accurate results for the Ising and Potts models in two dimensions, the convergence  of estimates for the problem of rods, where the correlation functions show oscillations, is  slow and the technique seems less reliable.  It was concluded that the second transition is continuous and not in the Ising universality class. We note that  this technique also indicates that the first transition from isotropic to nematic is not in the Ising universality class, contrary to strong existing evidence from other methods.

In three dimensions, there is no phase transition for $k \leq 4$. For $k\geq 7$, the system undergoes phase transitions from disordered to nematic to a layered disordered phase as density is increased. In the layered disordered phase, the system breaks up into very weakly interacting two dimensional planes within which the rods are disordered. For $4 < k< 7$, there is no nematic phase, and a single phase transition from a disordered to a layered disordered phase~\cite{2017-vdr-jsm-different,2017-gkao-pre-isotropic}. The nematic to layered disordered phase is expected to be similar to that in two dimensions. However, it is difficult to numerically  study  this transition because, in finite systems, these two phases sandwich a third thermodynamically unstable layered nematic phase~\cite{2017-vdr-jsm-different}.
 
It is thus clear that, in spite of several studies, the transition from nematic to HDD phase as well as the nature of the HDD phase are  poorly understood. Current numerical evidence suggests a continuous phase transition with the universality class being ambiguous. The only established results for the high density phase is for the fully packed phase. For this special point, it is known that the correlations between orientation of rods decrease algebraically with distance~\cite{1961-k-physica-statistics,kasteleyn1963dimer,1963-fs-pr-statistical,2007-gdj-pre-random,kenyon2000conformal,2003-hkms-prl-coulomb}. Also,  it has been conjectured that  the entropy per site in the full packing problem, on $d$-dimensional hypercubical lattices,  shows  hyper-universal behavior in the limit of  large $k$:  the leading term  is $A (\ln k)/ k^2$, with  the coefficient $A =1$, independent of $d$~\cite{dhar2021entropy}.

In this paper, we argue that the  phase transition from the nematic to the HDD phase is a first order transition. For large $k$,  the value of the critical fugacity $z^*(k)$ at this transition is shown to be $[ k/\ln k]^k$ to leading order in $k$. The density of holes is shown to jump from a value exponentially small in $k$ to a value $\ln k/k^2$ to leading order in $k$. We present strong, but not rigorous arguments, based on perturbation theory, that our results are asymptotically exact for large $k$.  These results are consistent with an exact solution that we obtain for a strip of size $k \times \infty$.  We finally present Monte Carlo evidence for the presence of a first order transition for $k\geq 9$.

The remainder of the paper is organized as follows. In Sec.~\ref{sec:model}, we define the model precisely. In Sec.~\ref{sec:nematicphase}, we recapitulate the results of the one dimensional problem that will be used later. We also discuss the  perturbation theory about the fully ordered nematic state at arbitrary densities, and show that for large $k$, for most of the density values in the nematic range, the deviations of various properties from the fully ordered nematic state are negligibly small.  
In Sec.~\ref{sec:tangent}, using only the fact that the fully packed state has a finite entropy per site, we show that at 
high densities, it is entropically favorable for the system in the nematic state with uniform density to phase separate into two phases, one with full packing, and the other nematically ordered at lower density.  We use this fact to estimate the density beyond which the instability sets in, and the corresponding chemical potential. In Sec.~\ref{sec:holes}, we define two approximations for the HDD phase called HDD$_1$ and HDD$_2$ phases,  which includes vacancies and allows for exact calculation of the partition function. We verify that this improved calculation of entropy does not change the basic conclusions of  Sec.~\ref{sec:tangent}. 
In Sec.~\ref{sec:strip}, we discuss a technique to determine the exact partition function per site for $k$-mers on a strip of width $k$. We cannot obtain a closed form solution, but instead devise an algorithm to determine numerically  the asymptotic value  of partition function per site in the thermodynamic limit, for a given numerical value of the  rod activity $z$. The method involves summing a series numerically. The convergence  is somewhat nontrivial, but  we are able to  determine the density of covered sites as a function of activity $z$ to about $8$ digit accuracy for each value of $z$. This one dimensional problem does not show a strict phase transition, but has a very sharp increase in the density  near   $z^*$.  The value of $z^*$ and the nearly sharp jump in density can be determined, and agree with the conclusions of the simpler calculations in Secs.~\ref{sec:tangent} and \ref{sec:holes}.  Section~\ref{sec:Montecarlo} contains the results of fixed density Monte Carlo simulations. We present  strong evidence of two phase coexistence  for $k=9$, which  is a signature of a first order transition. Finally, Sec.~\ref{sec:conclusions} contains a summary of our results and some concluding remarks. 

\section{\label{sec:model} definition of the Model}

In this section, we define the model more precisely and set the notation. 

Consider a $ L \times L$ square lattice,  with open  boundary conditions. A rod or $k$-mer occupies $k$ consecutive sites along one of the $x$- or $y$-directions. A site can  have at most one $k$-mer passing through it. We will consider mono-dispersed systems of $k$-mers. The weight or activity of each $k$-mer is $z=e^{\mu}$, where $\mu$ is the reduced chemical potential, and we  have set the inverse temperature $\beta =1$.

We refer to rods pointing in the $x$- and $y$-directions as $x$-mers and  $y$-mers respectively. The density $\rho$ will denote the fraction of sites covered by $k$-mers, while the density of vacant sites will be denoted by 
\be
\epsilon = 1-\rho.
\ee 
We will denote  the fraction of sites covered by $x$-mers and $y$-mers as $\rho_x$ and $\rho_y$ respectively.  The nematic order parameter $Q$ is defined as 
\be
Q= \frac{\left| \rho_x - \rho_y \right|}{\rho},
\ee
with $Q$ being zero in the disordered phase and one in the perfectly ordered nematic phase.

We will denote by $S_{d,k}(\epsilon)$ the entropy of the system of $k$-mers in $d$ dimensions at hole density $\epsilon$. Since the values of $d$ and $k$ are fixed in most of our
discussion, we will suppress these indices if the meaning is clear by the context.

\section{\label{sec:nematicphase} Description of the nematic phase}

Our key observation in this work is that in the nematic phase, for large $k$, the nematic order is very strong, with deviations of the  order parameter $Q$ from the maximum value of $1$ being negligible  in most of the range of the nematic phase. This may be seen as follows. Consider a generalized problem with different activities $z$ and $z_y$ for the $x$-mers, and $y$-mers.  Let the corresponding grand partition function for an $L \times L$ lattice by denoted by $\Omega_{2d}( L, z, z_y)$. If $z_y =0$, then the partition function factorizes into partition function of $L$ one-dimensional chains:
\begin{equation}
\Omega_{2d}( L, z, 0) = \left[ \Omega_{1d}(L, z) \right]^L,
\end{equation}
where $\Omega_{1d}( L,z)$ is the grand partition function for a  open linear chain of $L$ sites.  It satisfies the recursion relation
\begin{equation}
\Omega_{1d}( L, z)  = \Omega_{1d}( L-1, z) + z \Omega_{1d}(L-k, z).
\label{eq:recursion}
\end{equation}
For large $L$, $\Omega_{1d}( L,z) \sim \lambda^L$, where $\lambda$ is the solution of the algebraic equation
\begin{equation}
\lambda^k - \lambda^{k-1} = z.
\label{eq:lambda}
\end{equation}

The density of covered sites is obtained from the partition function by differentiating $\lambda$  with respect to $z$: $\rho=k z d(\ln \lambda)/dz$. It is easily verified that one obtains a rather simple result
\begin{equation}
\frac{\epsilon}{1 - \epsilon} = \frac{1}{ k(\lambda -1)}.
\label{eq:eps}
\end{equation}
We can also obtain the entropy per site for this fully nematic state, to be denoted by $S_{nem}(\epsilon)$, as a function of the density of holes $\epsilon$. The enumeration reduces to the arrangement of $L(1-\epsilon)/k$ rods and $L \epsilon$ holes. Thus,
\begin{align}
&S_{nem}=\lim_{L\to \infty} \frac{1}{L} \ln \binom{L\left(\frac{1-\epsilon}{k}+\epsilon\right)}{\frac{L (1-\epsilon)}{k}}, \nonumber \\
&= \left[\frac{1 -\epsilon}{k} + \epsilon \right]  \ln \left[ \frac{1 -\epsilon}{k} +\epsilon \right] 
-\frac{1 -\epsilon}{k} \ln \left[ \frac{1 -\epsilon}{k}\right]  - \epsilon \ln \epsilon. 
\label{eq:snem}
\end{align}

To include $y$-mers in this perfectly ordered phase, we expand  $\Omega_{2d}( L, z,z_y)$ as a power series in $z_y$, and put $z_y =z$ at the end of the calculation. Expanding to  linear order in $z_y$,  we obtain
\be
\frac{\Omega_{2d}( L, z,z_y)}{\Omega_{2d}( L, z,0)}=   1+ L^2 z_y \left[\frac{\Omega^\prime_{1d}(L, z,0)}{\Omega_{1d}(L, z,0)}\right]^k + \mathcal{O}(z_y^2),
\label{eq:8}
\ee
where $\Omega^\prime_{1d}(L, z,0)$ is the partition function of a one dimensional system in which one fixed site is empty. This immediately implies that the second term on the right hand side of Eq.~(\ref{eq:8}) is the $k$th power of the hole density $\epsilon$. Thus,
\bea
\frac{1}{L^2}\ln \Omega_{2d}( L, z,z_y) &=& \ln \lambda +z_y  \epsilon^k + \mathcal{O}(z_y^2).
\eea

For moderately small values of $\epsilon$, and large $k$, $\epsilon^k$ becomes very small. However, as $\epsilon$ tends to zero, the coefficient $z_y$, when set equal to $z$ becomes large.  For $\epsilon \to 0$,  $\lambda$ varies as $1/ \epsilon$, and $z$ varies as $\lambda^k$, and hence as  $\epsilon^{-k}$ [see Eqs.~(\ref{eq:lambda}) and (\ref{eq:eps})]. Thus, for $\epsilon$ tending to zero, the term $z_y  \epsilon^k$ tends to a finite constant. 

\begin{figure}
\includegraphics[width=0.825\columnwidth]{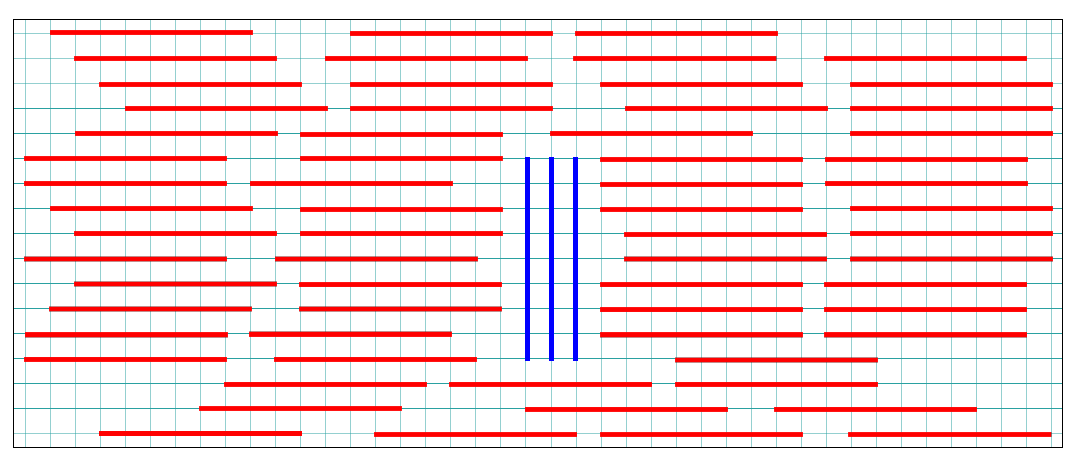}
\caption{\label{fig:singlecluster}  (Color online)  A cluster of $r=3$ vertically aligned $y$-mers  in a sea of $x$-mers.}
\end{figure}

It turns out  that in the limit of small $\epsilon$, the term corresponding to $r$ parallel vertical rods (see Fig.~\ref{fig:singlecluster} for an example) also contributes to order $\epsilon^k$. Hence, it is desirable to sum over the such configurations, and find  the total weight of placing $r$  adjacent parallel aligned vertical rods, summing over $r$. 

A similar calculation to the one given above for the contribution to the free energy from a single $y$-mer gives  that the weight of a configuration with $r$ such $y$-mers in a sea of $x$-mers  is $\epsilon^k z^k \lambda^{-(r-1)k} $. Summing over this geometric series, we obtain the net contribution  of these configurations in the series expansion for $ \ln \Omega_{2d}( L, z,z_y)$ as
\be
\frac{1}{L^2}\ln \Omega_{2d}( L, z,z_y) \approx \ln\lambda + F,
\ee
where 
\be
F =  z \lambda \epsilon^k= \left[\frac{1+ (k-1)\epsilon}{k} \right]^k  \frac{1-\epsilon}{\epsilon k}.
\label{eq:F}
\ee
When $\epsilon$ tends to zero, $F$ diverges as $1/\epsilon$. Thus, while the contribution from a single $y$-mer  ($r=1$) tended to a constant for $\epsilon$ tending to zero, the contribution from the sum of islands diverges for small $\epsilon$. This analysis shows that for  very small $\epsilon$, the purely nematic state is unstable to nucleation of stacks of vertical rods, signaling the onset of the HDD  phase.

\begin{figure}
\includegraphics[width=\columnwidth]{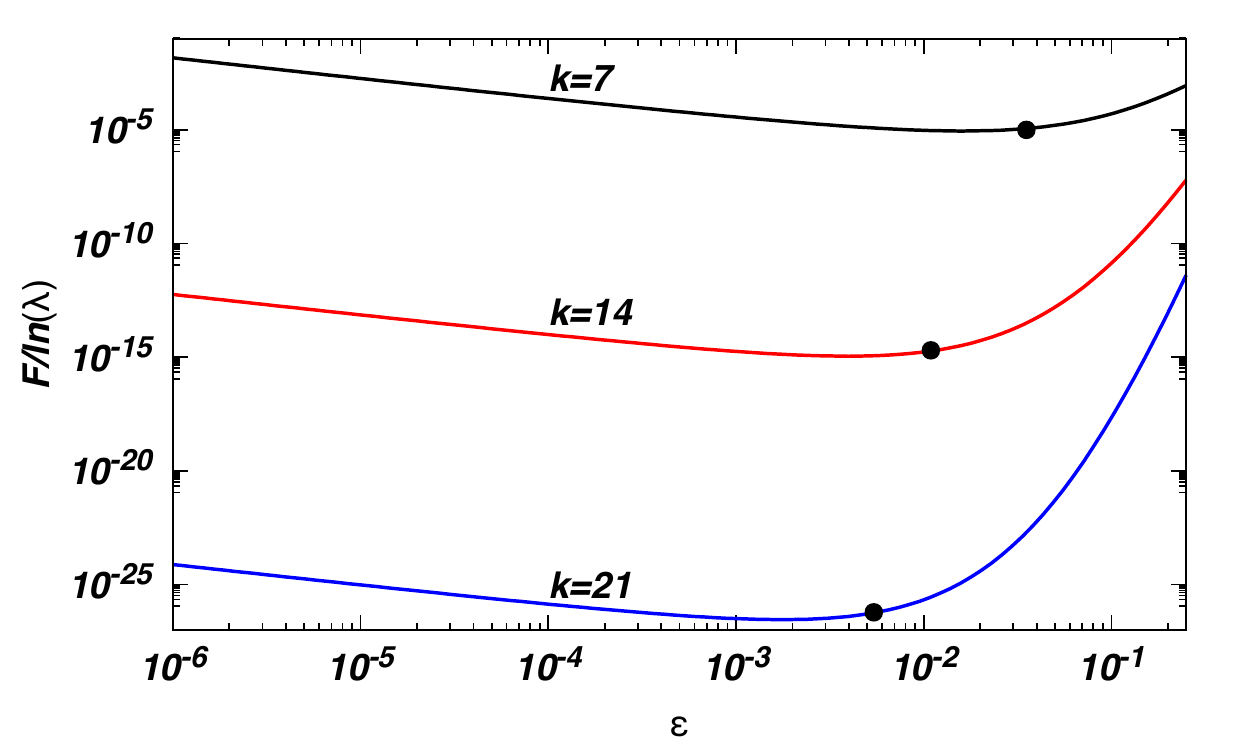}
\caption{\label{fig:Fepsilon} (Color online) The variation of the relative contribution of the correction to entropy due to the presence of islands [see the factor $F$ in Eq.~(\ref{eq:F})] with $\epsilon$ for different values of $k$. The solid black circles on each curve is at the value of $\epsilon$ at which the nematic phase becomes unstable, as estimated from the tangent construction in Sec.~\ref{sec:tangent}.}
\end{figure}

In Fig.~\ref{fig:Fepsilon}, we show the variation of the relative contribution $F/\ln \lambda$ with $\epsilon$. The first order correction to the nematic state  is negligibly small, especially for large $k$. In Fig.~\ref{fig:Fepsilon}, the solid black circles denote the value of $\epsilon$ at which the pure nematic phase becomes unstable to a two-phase coexistence regime, as estimated by the tangent construction calculation in  Sec.~\ref{sec:tangent}. At the approximate transition point, the relative correction $F/\ln \lambda$ is approximately equal to $10^{-5}$, $2 \times 10^{-15}$, $6 \times 10^{-27}$ for $k=7,14,21$ respectively. Thus, for large $k$, for a substantial  range  of densities for which the phase is nematic, 
\be
S(\epsilon) \approx S_{nem}(\epsilon) + \mathcal{O} (k^{-k}),
\label{eq:approximation}
\ee
where we have dropped the subscripts of $S_{2, k}(\epsilon)$ which denotes the true entropy per site of the full two-dimensional problem.

\section{\label{sec:tangent}The tangent construction} 

The entropy $S(\epsilon)$ is a convex function of $\epsilon$ which implies that a tangent drawn at any point lies above the curve, and in particular is larger than the entropy at full packing:
\begin{equation}
S(\epsilon) - \epsilon \frac{d}{d\epsilon} S(\epsilon) \geq S(0).
\label{inequal}
\end{equation}

As discussed in Eq.~(\ref{eq:approximation}), $S(\epsilon)$ is approximated very well  by $S_{nem}(\epsilon)$ in the nematic phase.
From Eq.~(\ref{eq:snem}), we note  that $S_{nem}(\epsilon)$ tends to zero when $\epsilon$ tends to zero. But, we know that $S(0)$, the entropy of the the fully packed  phase, is nonzero and  varies as $k^{-2} \ln k$ for large $k$~\cite{dhar2021entropy}.

\begin{figure}
\includegraphics[width=\columnwidth]{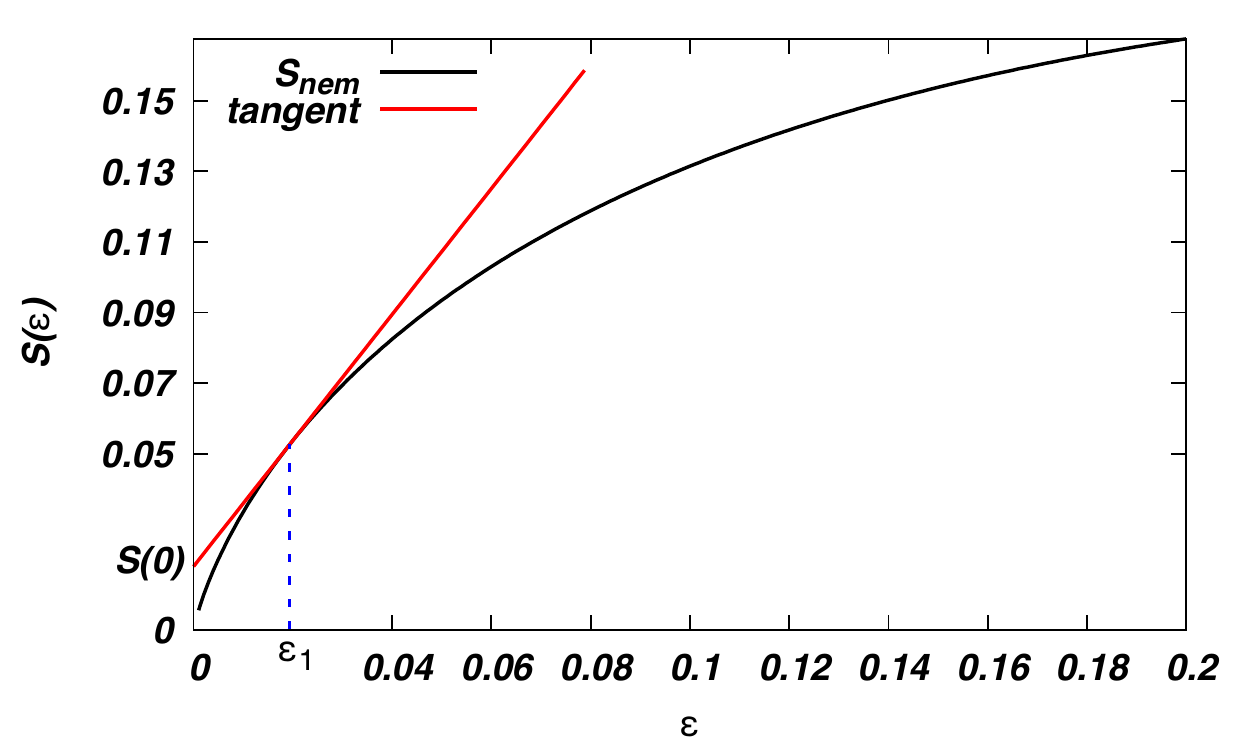}
\caption{\label{fig:tangentconstruction} (Color online) The construction of the tangent from $S(0)$ to the curve $S_{nem}(\epsilon)$ is shown for $k=10$. $S(0)$ is obtained from the solution of the fully packed $10 \times \infty$ strip. }
\end{figure}

Suppose we do not use any information about the behavior of the function $S(\epsilon)$ in the HDD phase, other than the fact that $S(0) >0$. Then, the expression for $S_{nem}(\epsilon)$ given in Eq. (\ref{eq:snem}) does not satisfy the inequality in Eq.~(\ref{inequal}) for small enough $\epsilon$. Suppose we draw a tangent to the curve $S_{nem}(\epsilon)$ from the point $(\epsilon=0, S(0))$. Let this tangent meet  the curve $S_{nem}(\epsilon)$ at $\epsilon=\epsilon_1$. Figure~\ref{fig:tangentconstruction} shows an example for $k=10$, where the fully packed entropy was approximated by its lower bound: the entropy for that of a strip $k \times \infty$.
This   tangent  would  be above the curve  $S_{nem}(\epsilon)$ in the range $ 0 < \epsilon < \epsilon_1$. This implies that for hole density less than $\epsilon_1$, it is entropically advantageous for the system to separate into two phases, one of  density $\epsilon_1$ and the other of density $1$ rather than have a phase with uniform hole density $\epsilon$.

From  Eqs.~(\ref{eq:snem}) and (\ref{inequal}), it is easily seen that the equation determining $\epsilon_1$ simplifies to 
\begin{equation}
\frac{1}{k} \ln \left[ \frac{ 1 +(k-1) \epsilon_1}{1 -\epsilon_1} \right] = S(0).
\label{eq:14}
\end{equation}
Using the fact that $S(0)\approx  (\ln k)/k^2$, for $k\gg 1$,  then for $k \epsilon_1 \ll 1$, Eq.~(\ref{eq:14}) simplifies to
\bea
\epsilon_1 &\approx& S(0), \label{eq:15} \\
&\approx& \frac{\ln k}{k^2}, ~{\rm for}~k \gg 1.
\label{eq:epslion1tangent}
\eea

The value of the chemical potential $\mu^*$  at this first order transition is related to the slope of the tangent:  $\mu^*=k dS_{nem}/d \epsilon|_{\epsilon_1}$. To leading order, $dS_{nem}/d \epsilon \approx -\ln(k \epsilon)$. Thus, we obtain $\mu^*$ to leading order in $k$ as
\bea
\mu^*  &\approx& -k\ln \left[ k S(0)\right], \label{eq:16} \\
&\approx&k \ln \left[ \frac {k}{ \ln{k}}\right], ~{\rm for}~k \gg 1.
\label{eq:mustartangent}
\eea

In view of Eq.~(\ref{eq:approximation}) being true in all dimensions, we expect that Eqs.~(\ref{eq:15}) and (\ref{eq:16}) are asymptotically exact in all dimensions, and we  are led to conjecture that  for hypercubical lattices in all dimensions $d \geq 2$, 
\bea
&&\lim_{ k \rightarrow \infty} \frac{ \epsilon_1(k)}{S_{d,k}(0)} =1,\\
&&\lim_{k \rightarrow \infty} \exp\left[ \frac{\mu^*(k)}{k} \right] k S_{d,k}(0)  =1,\\
&&\lim_{k \rightarrow \infty} \frac{\epsilon_2(k) k ^m}{\epsilon_1(k)} =0, ~{\rm for~ all}~m>0.
\eea
Here, for clarity,  in a departure from our notations used in the paper, we have explicitly displayed the $d$ and $k$ dependence of $\epsilon_1$, $\epsilon_2$, $\mu^*$ and $S_{d,k}(0)$.

We can check for the consistency of the assumption that even at $\epsilon =\epsilon_1$, the entropy in the nematic state is well approximated  by $S_{nem}(\epsilon_1)$, by noting that the value of the factor $F$ is of order $1/k^{k-1}$ is very small, even for $k =7$ (see Fig.~\ref{fig:Fepsilon}).

In Table~\ref{table:table}, we list the values of $\epsilon_1$ and $\mu^*$ for $k$ from $7$ to $21$, obtained from the tangent construction. For this calculation, we do not know the exact value of $S(0)$. Instead, we use a lower bound to $S(0)$, obtained by solving for the entropy of a fully packed $k\times \infty$ strip. Then $S(0) \geq \ln \phi$, where $\phi$ is  the solution of 
\be
\phi^k -\phi^{k-1}=1. 
\label{eq:phi}
\ee
We then set $S(0)\approx k^{-1} \ln \phi$. 
 \begin{table}
\caption{\label{table:table}The  density  of the nematic phase in the coexistence regime, $\epsilon_1$,  and the critical chemical potential $\mu^*$ for different $k$, as obtained from the tangent construction. }
\begin{ruledtabular}
\begin{tabular}{ccc}
$k$ & $\epsilon_1$ & $\mu^*$\\
\hline
$7$ & $ 0.0352044085666$ & $ 10.9186783149$ \\
$8$ & $ 0.0281891510465$ & $ 13.1470420927$ \\
$9$ & $ 0.0231353803495$ & $ 15.4576043505$ \\
$10$ & $ 0.0193666682476$ & $ 17.8426607700$ \\
$11$ & $ 0.0164763742454$ & $ 20.2958267516$ \\
$12$ & $ 0.0142077918145$ & $ 22.8117292904$ \\
$13$ & $ 0.0123921831043$ & $ 25.3857861041$ \\
$14$ & $ 0.0109148059700$ & $ 28.0140438975$ \\
$15$ & $ 0.00969534939667$ & $ 30.6930575308$ \\
$16$ & $ 0.00867618606129$ & $ 33.4197980784$ \\
$17$ & $ 0.00781506083686$ & $ 36.1915817038$ \\
$18$ & $ 0.00708039604466$ & $ 39.0060138133$ \\
$19$ & $ 0.00644819338532$ & $ 41.8609446150$ \\
$20$ & $ 0.00589993952465$ & $ 44.7544333257$ \\
$21$ & $ 0.00542115940916$ & $ 47.6847190234$ 
\end{tabular}
\end{ruledtabular}
\end{table}

\begin{figure}
\includegraphics[width=\columnwidth]{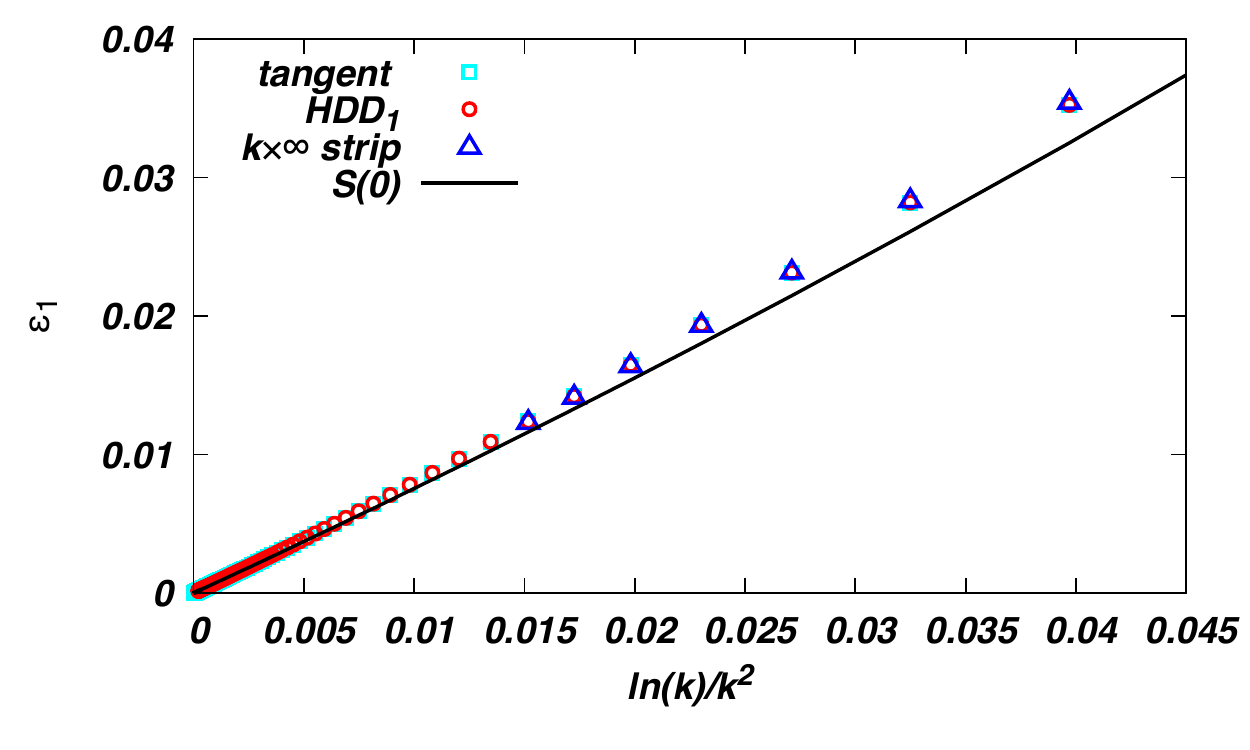}
\caption{\label{fig:epsilon1k}  (Color online) The variation of nematic coexistence density $\epsilon_1$ with $(\ln k)/k^2$, obtained from the three different calculations. The data points are for $k=7$ to $k=30$ (tangent), $k=150$ (HDD$_1$), and $k=13$ ($k\times \infty$ strip). The solid line is the estimate in Eq.~(\ref{eq:15}).  The HDD$_1$ phase is defined in Sec.~\ref{sec:holes} and the solution for the strip is given in Sec.~\ref{sec:strip}.}
\end{figure}

Figure~\ref{fig:epsilon1k} shows the variation of $\epsilon_1$ with 
$(\ln k)/ k^2$ for different $k$. The data is compared to the asymptotic answer $S(0)$ [see Eq.~(\ref{eq:15})]. We see that the asymptotic form is rather good even for fairly small values of $k$. 

\begin{figure}
\includegraphics[width=\columnwidth]{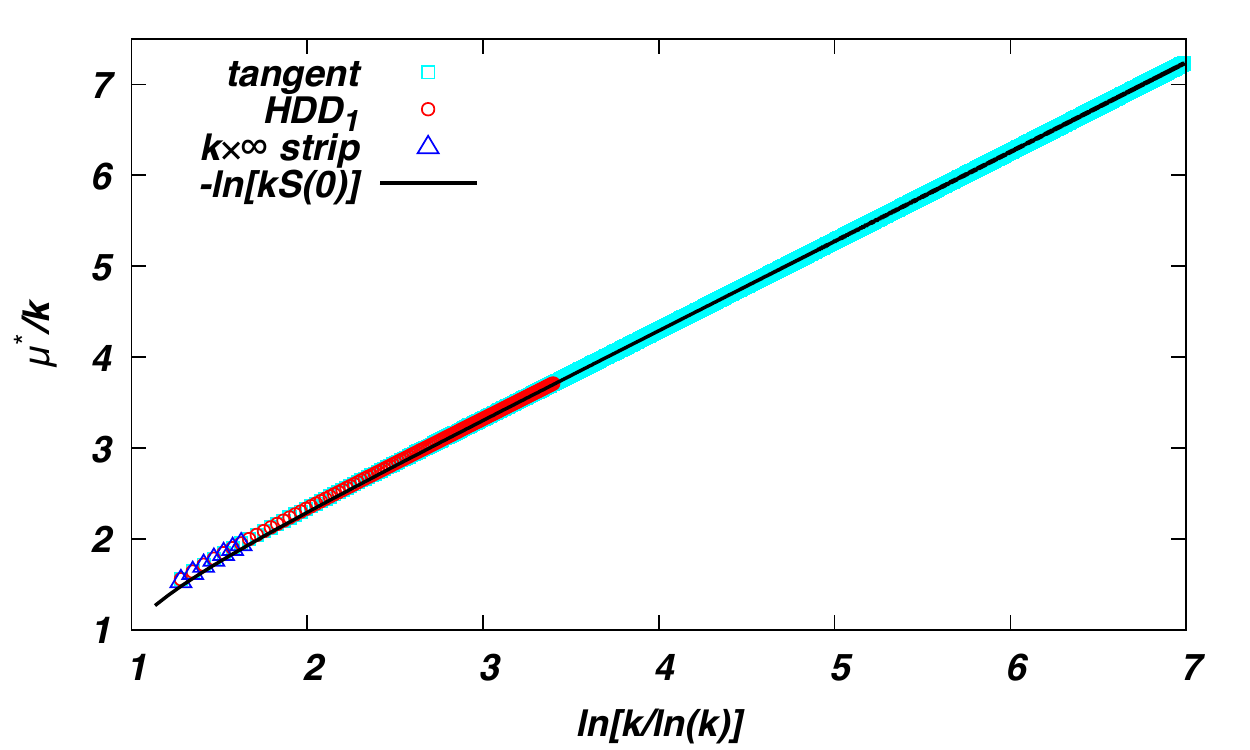}
\caption{\label{fig:mustark}  (Color online) The variation of scaled critical chemical potential $\mu^*/k$  with $\ln(k/\ln k)$, obtained from the three different calculations. The data points are for $k=7$ to $k=30$ (tangent), $k=150$ (HDD$_1$), and $k=13$ ($k\times \infty$ strip). The solid line is the estimate in Eq.~(\ref{eq:16}). The HDD$_1$ phase is defined in Sec.~\ref{sec:holes} and the solution for the strip is given in Sec.~\ref{sec:strip}.  }
\end{figure}

Figure.~\ref{fig:mustark}, shows the variation of  $ \mu^*/k$ with $ \ln [k/\ln k]$. Again, we see a fair agreement with the asymptotic expression as given in Eq.~(\ref{eq:16}). 

\section{\label{sec:holes}Taking into account the effect of holes  in the HDD phase}

In the analysis of Sec.~\ref{sec:tangent},  the HDD phase was approximated as fully packed with $\epsilon=0$. This, of course, cannot be correct, as at any finite chemical potential, there will be a finite density of holes in the HDD phase as well. We will now show that even when holes are accounted for in an approximate way,  the basic features of the simple calculation presented in Sec.~\ref{sec:tangent} are still preserved.

\begin{figure}
\includegraphics[width=\columnwidth]{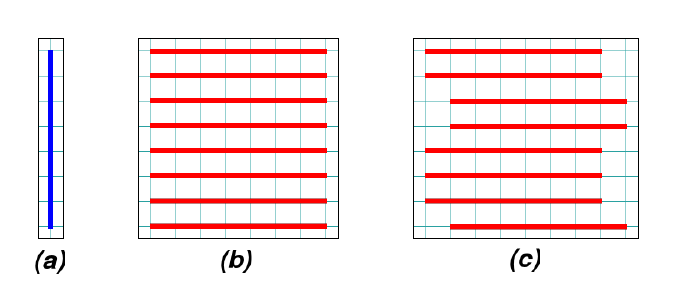}
\caption{\label{fig:basicconfig}  (Color online) The three basic tiles used in the definition  the HDD$_1$ phase are shown for $k=5$. (a) A $y$-mer, (b)  a $k \times k$  square covered by $k$ $x$-mers,  and (c) a $(k+1) \times k$ tile with $k$  $x$-mers, each of which can be in two possible positions.}
\end{figure}

Since we are not able to exactly calculate the entropy of the  HDD phase, we will approximate it by a reference phase,  to be called HDD$_1$ phase. This   shares some qualitative features with the actual HDD phase, but allows us to calculate the corresponding partition function exactly. In the HDD$_1$ phase, we impose some  restrictions on the  configurations accessible to the system (just as we did for the nematic phase). In the HDD$_1$ phase, the system is made of $L \times k$ strips, with no $k$-mers shared between different strips.  In each strip, the configuration is made by concatenating  copies of the  three basic patterns, called tiles,   shown in Fig.~\ref{fig:basicconfig}: (1) A $1 \times k$ tile consisting of  a $y$-mer, (2)  a $k \times k$  square tile covered by $k$ $x$-mers, and (3) a $(k+1) \times k$ tile, with $k$  $x$-mers each of which can be in two possible positions.  Thus, there are  $2^k$ distinct tiles of the third type. 
The total sum of weights of these three types of tiles are  $x z$, $x^k z^k$ and $(2 z)^k x^{k+1}$ respectively,  where the power of $x$ is the horizontal  extent of  the tile.

Lest the construction appear very contrived, we note that the first two tiles already give a nonzero entropy per site in the full packing limit, and the entropy varies as $(\ln k)/k^2$ for large $k$~\cite{dhar2021entropy}. Thus this reproduces the exact asymptotic  behavior of $S_{2d}(0)$, for large $k$. The third tile allows for vacancies. We can start with a fully packed $(k + 1)  \times k $ rectangle with one $y$-mer and $k$ $x$-mers, and remove the $y$-mer. Then the adjacent $x$-mers  can now move by one lattice site independent of each other. Thus the total number of allowed configurations is $2^k$, for each rod removed. Thus, we  retain an important feature of the $k$-mer problem: the number of new configurations are exponentially large in $k$ for each $k$-mer that is  removed.

Let the grand partition function of an $L \times k$  lattice in the HDD$_1$ phase be $\Omega_{\mathrm{HDD}_{1}}(L,z)$. We define the generating  function
\begin{equation}
\Omega_{{\rm HDD}_1}(x,z) = \sum_{L=0}^{\infty}  x^L \Omega_{{\rm HDD}_{1}}(L,z).
\end{equation}
 We define $R(x,z)$ as  the sum of the weights of the constituent tiles
\begin{equation}
R(x,z) = xz + x^k z^k +  2^k  x^{k+1} z^k. 
\end{equation}
It is then easily seen that 
\bea
\Omega_{{\rm HDD}_1}(x,z) &=&  1 + R(x,z) + R(x,z)^2 + R(x,z)^3 + \ldots, \nonumber\\
&=& \frac{1}{1 - R(x,z)}.
\eea

If the partition function per site in the HDD$_1$ phase is $\lambda_{{\rm HDD}_1}$,  then $ \lambda_{{\rm HDD}_1} =x^{*-1/k}$, where $x^*$ is the solution to the equation 
\begin{equation}
1=R(x^*,z) =  x^* z + x^{*k} z^k +  2^k x^{*k+1} z^k.
\end{equation}

\begin{figure}
\includegraphics[width=\columnwidth]{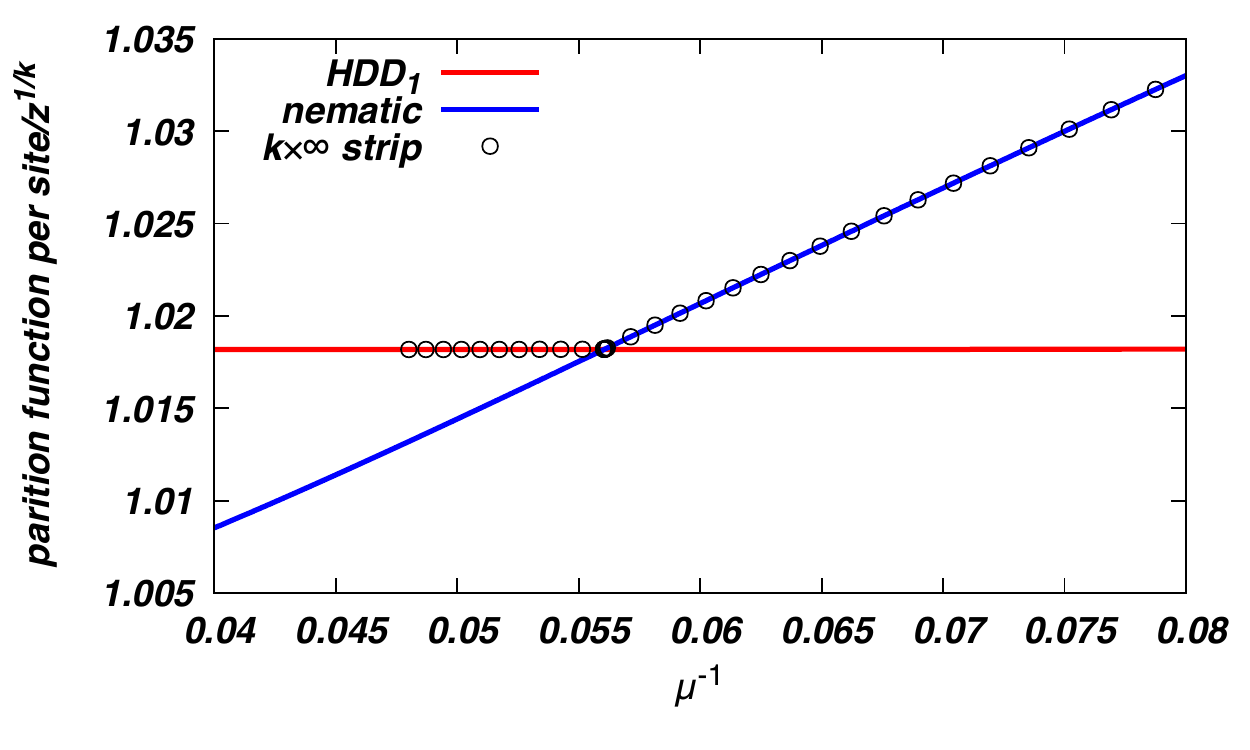}
\caption{\label{fig:lambda_mu_comp}  (Color online)  The variation of the scaled partition functions per site $\lambda_{{\rm HDD}_1}$ and $\lambda$ with chemical potential $\mu$ for $k=10$. The two curves intersect at  $\mu^*$. The partition function of the $k\times \infty$ strip is approximately equal to the largest of the two partition functions.}
\end{figure}

In Fig.~\ref{fig:lambda_mu_comp} we have plotted  the two approximate expressions for the partition function per site as function of the chemical potential for rods $\mu$ in the two phases.  These two curves $\lambda$ and $\lambda_{{\rm HDD}_1}$ intersect at at some point $\mu^*$. For $\mu < \mu^*$, the nematic phase has higher entropy, but for $\mu > \mu^*$, HDD$_1$ phase has a higher value of the partition function. Thus, the intersection of the two curves determines the  location of the first order transition. The discontinuity of the slope is related to the jump in density at the transition. 

The  equations determining $\mu^*=\ln z^*$ are
\be
\begin{split}
&\lambda^{*k}  - \lambda^{*k-1}=z^*, \\
&x^*= \lambda^{*-k}\\
&z^* x^* + (z^* x^*)^k  + (z^* x^*)^k 2^k x^* =1.
\end{split}
\label{eq:solution1}
\ee
These are easily solved for large $k$. 
For large $k$, Eq.~(\ref{eq:solution1}) has the  solution $ z^* x^* + (z^* x^*)^k  \approx 1$, with $ 2^k x^* \ll 1$.  Comparing with Eq.~(\ref{eq:phi}), we obtain $x^*z^*=\phi^{-1}$, where $\phi = \exp[k S(0)]$. On the other hand, from Eq.~(\ref{eq:solution1}), $x^*z^*= 1-\lambda^{*-1}$. Substituting for $x^*z^*$, we obtain
\begin{equation}
\lambda^* =  \frac{1}{ k S(0)}.
\label{eq:lambdastar}
\end{equation}


All the other critical parameters can now be calculated. From Eq.~(\ref{eq:eps}),   
\be
\epsilon_1\vert_{{\rm HDD}_1}\approx S(0) = \frac{\ln k}{k^2} ~ {\rm for}~ k \gg 1.
\ee
Also,
\be
\mu^*\vert_{{\rm HDD}_1}=\ln z^* \approx k \ln \frac{1}{ k S(0)} = k \ln \frac{k}{ \ln k} ~ {\rm for} ~k \gg 1.
\ee
We thus see that within HDD$_1$, though holes are taken into account, $\mu^*$ and $\epsilon_1$ do not change to leading order in $k$, when compared to the results obtained from tangent construction [see Eqs.~(\ref{eq:epslion1tangent}) and (\ref{eq:mustartangent})]. 

\begin{figure}
\includegraphics[width=\columnwidth]{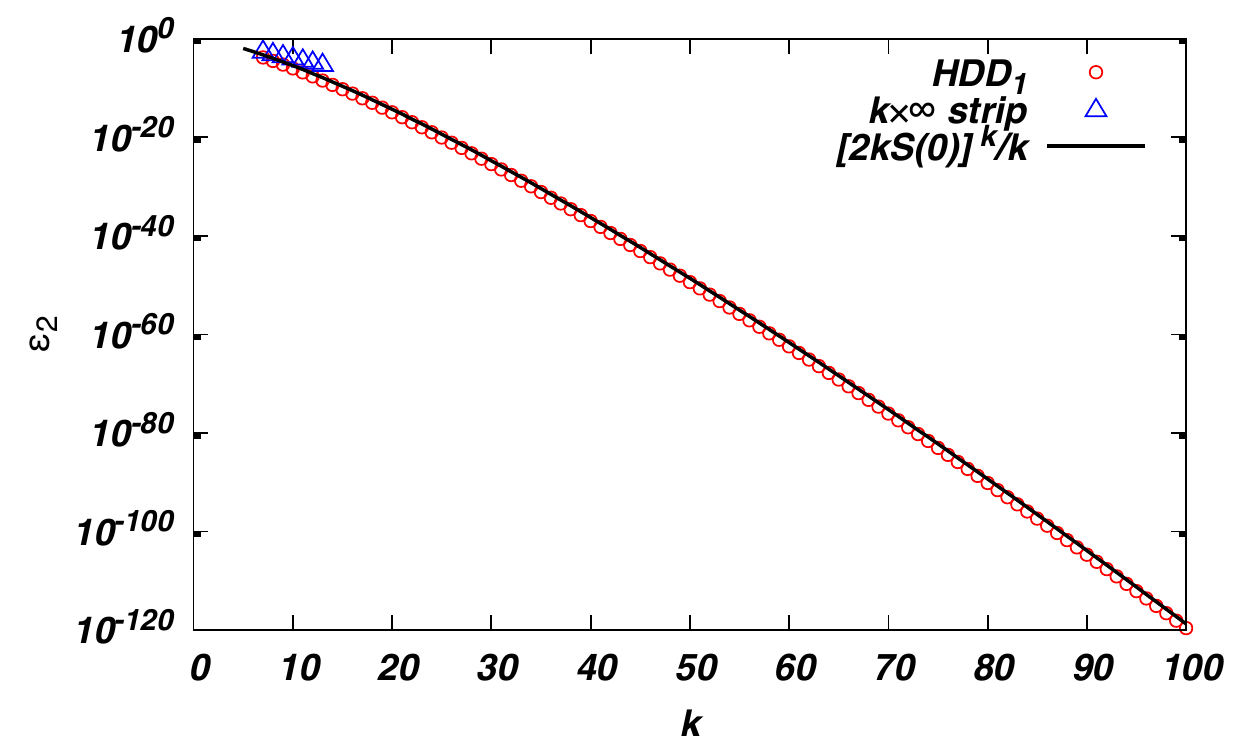}
\caption{\label{fig:epsilon2k}  (Color online) 
The variation of the HDD$_1$ coexistence density $\epsilon_2$ with
$k$, obtained from two different calculations. The data points are for 
$k=7$ to $k=100$ (HDD$_1$), and $k=13$ ($k\times \infty$ strip). The solid line is the asymptotic behavior  in Eq.~(\ref{eq:30}). 
}
\end{figure}

We now compute $\epsilon_2$, the hole density at the high density end of the coexistence region. Unlike in the tangent construction where $\epsilon_2=0$, in the HDD$_1$ phase, we obtain a non-zero answer for $\epsilon_2$. $\epsilon_2$ can be calculated from $x^*$ through $1-\epsilon_2= - k z^* d/dz^* (\ln x^{*-1/k})$. Simplifying, we obtain
\bea
&&\epsilon_2\vert_{{\rm HDD}_1}=\\
&&\frac{ 2^k x^* (1-\lambda^{*-1})^k}{1-\lambda^{*-1}+k (1-\lambda^{*-1})^k +2^k x^* (k+1) (1-\lambda^{*-1})^k}. \nonumber
\eea
For large $k$, substituting for $\lambda^*$ from Eq.~(\ref{eq:lambdastar}), we obtain
\bea
\epsilon_2\vert_{{\rm HDD}_1} &\approx& \frac{[2 k S(0)]^k}{k}, \label{eq:30}\\
&\approx & \frac{1}{k} \left[ \frac{2 \ln k}{k} \right]^k, ~{\rm for}~ k \gg 1.
\eea
Though $\epsilon_2$ is non-zero, it is smaller than exponential in $k$. Figure~\ref{fig:epsilon2k} shows the variation of $\epsilon_2$ with $k$. Comparing the  exact answer with the asymptotic result in Eq.~(\ref{eq:30}), we see a good match even for small $k$.

\begin{figure}
\includegraphics[width=\columnwidth]{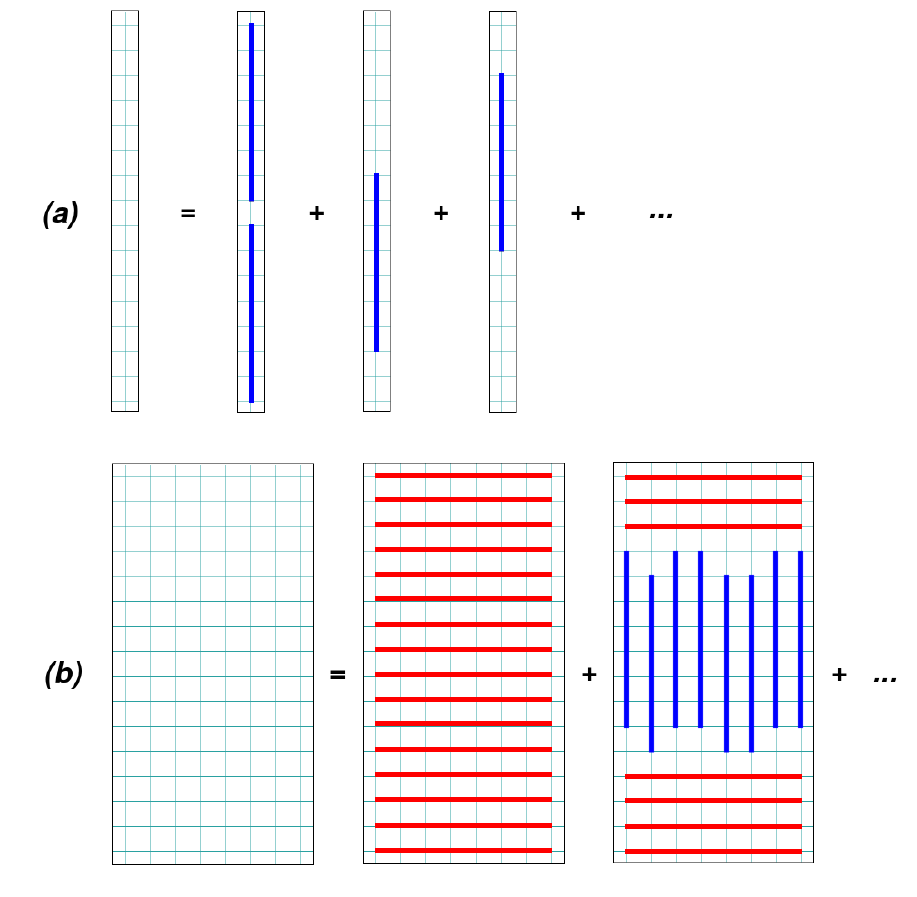}
\caption{\label{fig:basictiles}  (Color online) The two basic tiles used in the definition  the HDD$_2$ phase are shown for $k=5$. (a) A $1\times 2 k$ tile, and (b)  a $k \times 2k$  tile which contains at least one $x$-mer. }
\end{figure}

It is straightforward to make  improvements to the HDD$_1$  approximation. We outline the calculation using an example with the main conclusion being that the asymptotic results for the critical parameters do not change. In this example, we define a phase called HDD$_2$, in which  we break the lattice into  strips of width $2 k$. We use two types of tiles, as shown in Fig.~\ref{fig:basictiles},  to fill this strip.   The first is a $1 \times 2k$ tile that may have $0$ or $1$ or two $y$-mers. The combined weight of these tiles is $w_1(z) x$, where $w_1(z) = 1 + (k+1) z + z^2$.  The second type of tiles is of size $k \times 2k$. This may be filled in any way, by $x$-mers and $y$-mers
(all lying completely inside the tile), the only constraint being that there has to be at least one $x$-mer in the tile. This constraint ensures that concatenation of these two types of tiles has a unique decomposition into constituent tiles. 
The total weight of the second group of tiles is denoted by $w_2(z) x^k$, where $w_2(z)$ is a polynomial in $z$ with leading term being  $(k+2) z^{2k}$.

The  generating function for the width $2 k$ strip  in the HDD$_2$ phase is then
\be
\Omega_{{\rm HDD}_2} (x, z) = \frac{1}{ 1 -R_{{\rm HDD}_2}(x,z)},
\ee
where
\be
R(x, z)_{{\rm HDD}_2} = x w_1(z) + x^k w_2(z). 
\ee 
Let $x^*$ satisfy $R_{{\rm HDD}_2}( x^*,z)=1$. Then the partition function per site is $x^{*-[1/(2 k)]}$. Close to full packing, $(x^*z)$ tends to constant value $\alpha$. Taking only the leading power of $z$ in $w_2$, which is $(k+2) z^{2k}$, we obtain that $\alpha$ satisfies $\alpha + ( k+2) \alpha^k =1$. It is easy to  verify that this gives a higher entropy per site at full packing than that for the $k\times \infty$. However, the leading order contribution is the same and at the next order in $k$, the fractional correction is again of order $2^k/z$. The detailed analysis is similar to that for the HDD$_1$ phase, and we omit it  here.

\section{\label{sec:strip} Exact Solution of $k\times \infty$ strip}

In this section, we solve exactly for the entropy as well as the dependence of density and nematic order parameter on  chemical potential for the system of rods on a strip of size $k \times \infty$.

Let $\Omega_{strip}(L, z, z_y)$ be the partition function of an $L \times k$ strip, with the activities of the $x$-mers and $y$-mers being $z$ and $z_y$.  We define the generating function 
\be
\widetilde{\Omega}_{strip}(x, z, z_y)=\sum_{L=0}^\infty \Omega_{strip}(L, z, z_y) x^L.
\ee
\begin{figure}
\includegraphics[width=\columnwidth]{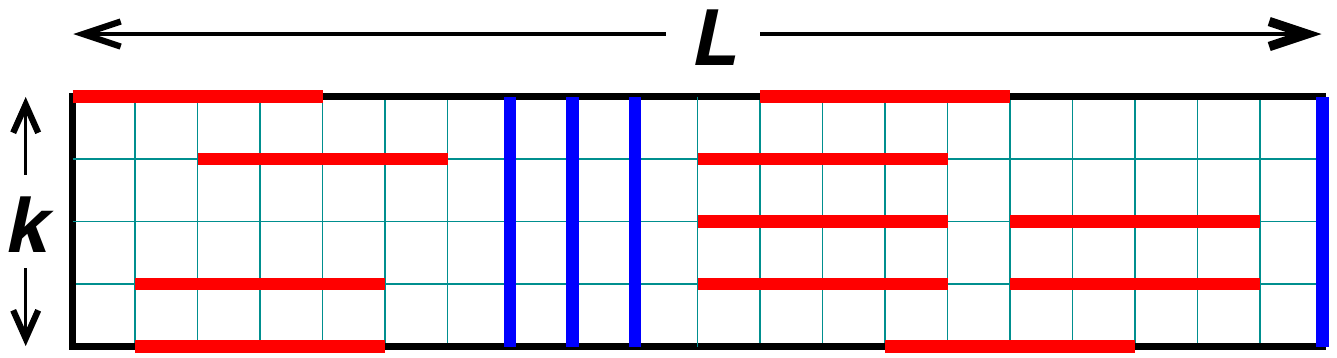}
\caption{\label{fig:stripconfig}  (Color online) An example of a configuration of rods in a $k\times L$ strip with $k=5$ and $L=21$. The $y$-mers, shown in blue, divides the strip into non-interacting segments containing only $x$-mers, shown in red.}
\end{figure}

Any given configuration of rods can be split into a segments of pure nematic phase of $x$-mers separated by $y$-mers, as illustrated in  Fig.~\ref{fig:stripconfig}. Given the positions of $y$-mers, each nematic segment can be filled independent of each other. $\widetilde{\Omega}_{strip}( x, z, z_y)$ can then expressed in terms of the generating function of the nematic segments. Let $\widetilde{R}(x, z_x)$ denote the generating function with only $x$-mers. Then,
\bea
\widetilde{\Omega}_{strip}(x, z, z_y)&=& \widetilde{R}(x, z) + \widetilde{R}(x, z) x z_y \widetilde{R}(x, z)\nonumber \\
&+& \widetilde{R}(x, z) x z_y \widetilde{R}(x, z) x z_y \widetilde{R}(x, z) + \ldots,
\label{eq:series} \\
&=& \frac{\widetilde{R}(x, z)}{1- x z_y \widetilde{R}(x, z)}.
\label{eq:series2}
\eea
The first to third terms in the right hand side of Eq.~(\ref{eq:series}) enumerate the  configurations with zero to two $y$-mers respectively, and so on.

The function  $\widetilde{R}(x, z)$ is easily expressed in terms of the one dimensional partition functions $\Omega_{1d}(L,z)$:
\be
\widetilde{R}(x, z) = \sum_{L=0}^\infty [\Omega_{1d}(L,z)]^k x^L.
\ee
The functions $\Omega_{1d}(L,z)$ increase exponentially with $L$, and would cause overflow problems in numerical evaluation of the series.  
To control this divergence, we eliminate the exponentially diverging part  by defining
\be
Q_L(z)= \lambda^{-L} \Omega_{1d}(L, z).
\ee
Here $\lambda$ is an implicit function of $z$ satisfying Eq.~(\ref{eq:lambda}). For these variables, the recursion relation 
Eq.~(\ref{eq:recursion}) reduces to
\be
Q_L(z)= \frac{Q_{L-1}(z)}{\lambda}+ \frac{\lambda-1}{\lambda} Q_{L-k}(z),~L\geq k,
\label{eq:q}
\ee
with the boundary conditions
\be
Q_L(z)=\lambda^{-L},~~L=0, 1, \ldots k-1.
\label{eq:qbc}
\ee
For large $L$, $Q_L(z)$ tends to  a finite value. 

Now, the partition function $\widetilde{\Omega}_{strip}(x, z, z_y)$ in Eq.~(\ref{eq:series2}) can be rewritten in terms of $Q_L(z)$ as
\be
\widetilde{\Omega}_{strip}(x, z, z_y)= \frac{\sum_{L=0}^{\infty}Q_L(z)^k \theta^L}
{1-\theta \frac{z_y}{z}(1-\lambda^{-1}) \sum_{L=0}^{\infty}Q_L(z)^k \theta^L},
\label{eq:omega-modified}
\ee
where 
\be
\theta=x \lambda^k.
\ee
The singularity $\theta^*$ closest to the origin of $\widetilde{\Omega}_{strip}(x, z, z_y)$ is given by the zero of the denominator in Eq.~(\ref{eq:omega-modified}):
\be
1-\theta^* \frac{z_y}{z}(1-\lambda^{-1}) \sum_{L=0}^{\infty}Q_L(z)^k \theta^{*L}=0.
\label{eq:45}
\ee
Knowing $\theta^*$ and hence $x^*$, we obtain the partition function to be 
\be
\lim_{L\to \infty} \frac{\ln \Omega_{strip}(L, z, z_y)}{L}=-\ln x^{*}.
\ee

While a closed form solution cannot be written down for $x^*$, it is possible to find a numerical solution.  For a given value of $z$,
we first find $\lambda$ using Eq~(\ref{eq:lambda}). To determine $Q_L$, we find note that for large $L$, $Q_L \to \lambda \epsilon$. We determine  the coefficients  $Q_L$ upto  $L=L^*$ till $Q_L^k -(\lambda \epsilon)^k< \Delta$ for $20$ consecutive $L$s.  We choose $\Delta=10^{-14}$.  The infinite sum in Eq.~(\ref{eq:omega-modified}) is split into a finite sum upto $L^*$ for $Q_L^k-(\lambda \epsilon)^k$ and an infinite sum over $(\lambda \epsilon)^k$.   We then determine $\theta^*$ using Eq~(\ref{eq:45}), and hence determine the partition function per site for the k-strip. The densities and nematic order parameter can be found by taking suitable numerical derivatives.

\begin{figure}
\includegraphics[width=\columnwidth]{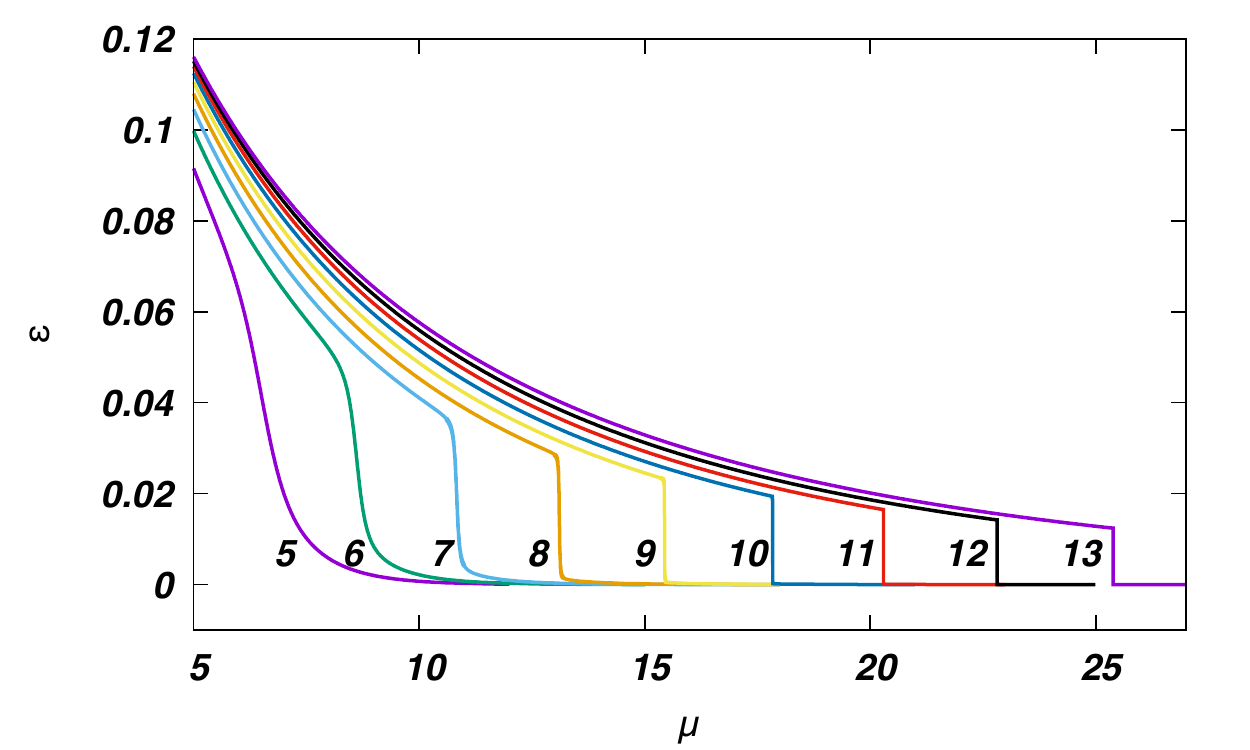}
\caption{\label{fig:murho}  (Color online) The variation of hole density $\epsilon$ with $\mu$ for different $k$, where the numbers in the plot refer to the value of $k$. The jump in density becomes sharper with increasing $k$. The data are obtained from the solution of the  $k\times \infty$ strip.}
\end{figure}

Figure~\ref{fig:murho} shows the variation of $\epsilon$ with $\mu$ for different $k$. The hole density $\epsilon$ shows a nearly discontinuous behavior,  which becomes sharper with increasing $k$. This jump occurs at larger $\mu$ and $\rho$ with $k$, as expected.

\begin{figure}
\includegraphics[width=\columnwidth]{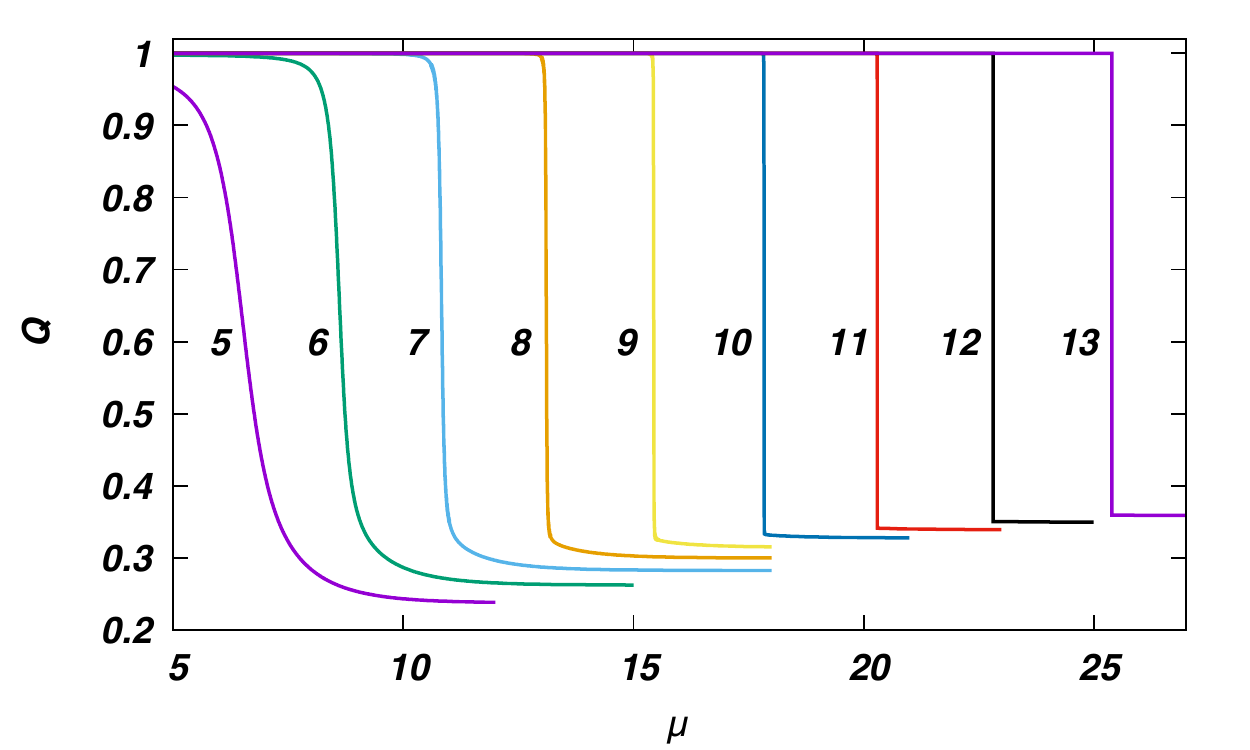}
\caption{\label{fig:munem}  (Color online) Variation of the order parameter $Q$ with $\mu$ for different $k$, where the numbers in the plot refer to the value of $k$. The jump in $Q$ becomes sharper with increasing $k$. The data are obtained from the solution of the  $k\times \infty$ strip.}
\end{figure}

More evidence for the near first order nature of the transition may be found by examining the order parameter $Q$. $Q$ shows a sharp decrease at small $\epsilon$, as shown in Fig.~\ref{fig:munem}. The discontinuity becomes sharper with increasing $k$. The nematic order parameter does not decrease to zero at full packing. This is an artifact of the finite width of the strip, and as the width increases $Q$ will be expected to decrease to zero in the entire HDD phase.

We estimate $\mu^*$ as the value of $\mu$ at which $d\rho/d\mu$ is maximum. It is determined to accuracy $10^{-4}$ for smaller $k$ and $10^{-5}$ for larger $k$. Figure~\ref{fig:mustark} compares the $\mu^*$ obtained from the strip with that of the tangent construction and HDD$_1$ phases. We observe that the three calculations give results that are not distinguishable.

We now study the dependence of $\epsilon_{1}$ and $\epsilon_2$, the coexistence densities, on $k$. We identify $\epsilon_{1}$ and $\epsilon_2$ as the lower and higher densities at the point of discontinuity. There is a certain ambiguity in measuring these critical densities because the solution on the strip has no true discontinuities. Figure~\ref{fig:epsilon1k} shows the variation of $\epsilon_{1}$ with $k$. As for $\mu^*$, the results obtained from the strip are indistinguishable from that obtained from tangent construction and HDD$_1$ phases.

Figure~\ref{fig:epsilon2k} shows the variation of $\epsilon_{2}$ with $k$.  The data obtained from the strip solution, unlike the data for $\epsilon_1$ and $\mu^*$, show slight discrepancy from the calculation based on HDD$_1$ phase. The data are however consistent with $\epsilon_2$ decreasing  exponentially with $k$ for large $k$.

We now quantify the deviation, $\delta S_{nem}$, of the entropy of the nematic phase, obtained from the solution of the strip, from the variational nematic entropy $S_{nem}$. In addition to checking whether this quantity is small in the nematic phase, we would also like to check how well our estimates for $\delta S_{nem}$, obtained from summing over all islands, quantify these corrections. The sum over single islands gave a correction term  $F$ [see Eq.~(\ref{eq:F})]. However, these calculations were for the two dimensional lattice. On the $k \times \infty$ strip, due to open boundary conditions, we have to divide $F$  by $k$ to account for lack of translational invariance. Thus, our estimates for the corrections are
\be
F_{strip}=\frac{F}{k}= \left[\frac{1+ (k-1)\epsilon}{k} \right]^k  \frac{1-\epsilon}{\epsilon k^2}.
\label{eq:Fstrip}
\ee

\begin{figure}
\includegraphics[width=\columnwidth]{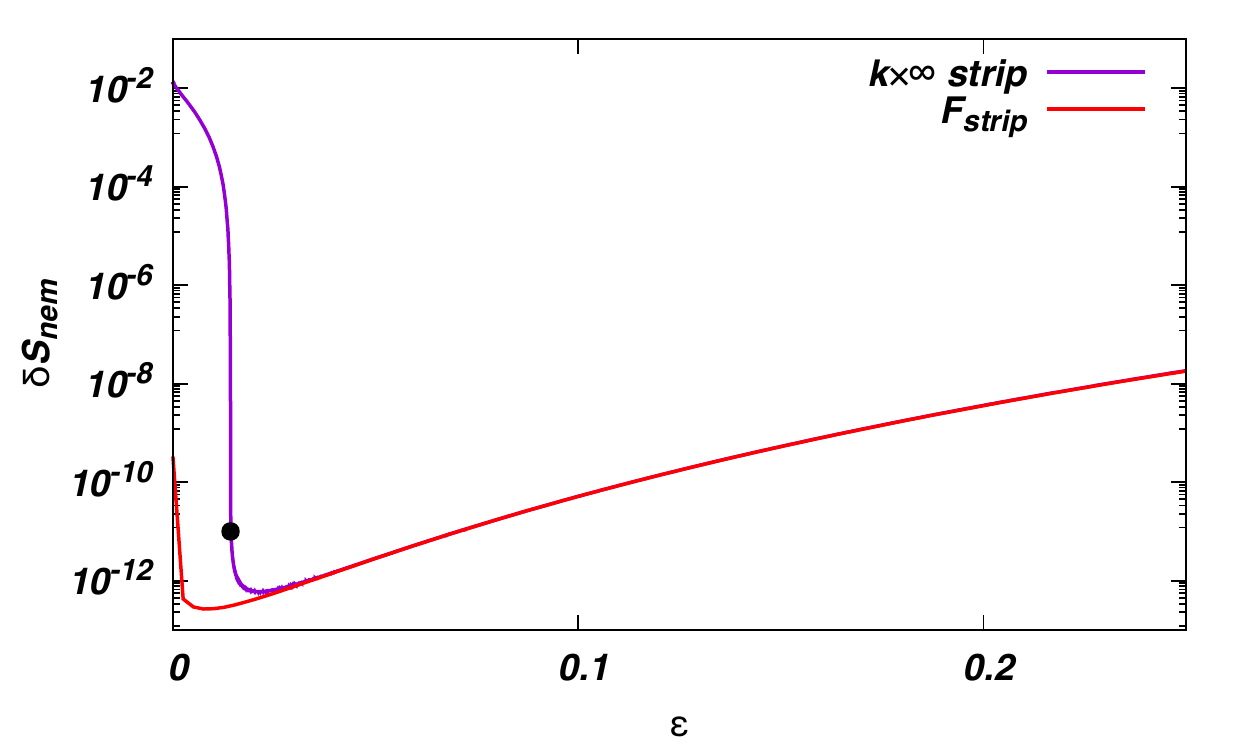}
\caption{\label{fig:entropycorrectionsk12}  (Color online) The deviation, $\delta S_{nem}$, of the entropy of the strip from $S_{nem}$ [see Eq.~(\ref{eq:snem})], as a function of hole density $\epsilon$ for $k=12$.  The data are compared with the theoretical estimates (denoted by $F_{strip}$) as given in Eq.~(\ref{eq:Fstrip}). The solid black circle is at $\epsilon_1$, as estimated from the solution of the system on the strip.}
\end{figure}

In Fig.~\ref{fig:entropycorrectionsk12}, we show the variation of $\delta S_{nem}$ with $\epsilon$ for $k=12$. First, we see that $\delta S_{nem}$ is very small in the nematic phase decreasing to as much as $10^{-12}$. There is a sharp increase in $\delta S_{nem}$ across the transition. Second, we compare the numerical result with our perturbative estimate from islands in Eq.~(\ref{eq:Fstrip}). As can be seen, the data match very well with Eq.~(\ref{eq:Fstrip}).

Figure~\ref{fig:lambda_mu_comp} compares the partition function per site  for the strip with $S_{nem}$ and that obtained from the HDD$_1$ phase for $k=10$.  The partition function of the strip  extrapolates between the two entropies. It also shows that the the variational estimate  $\lambda_{{\rm HDD}_1}$  for the high density phase describes the numerical solution of the strip quite well.

\section{\label{sec:Montecarlo} Monte Carlo simulations}

The arguments presented up to now  for the high density transition being discontinuous were for large $k$ or for a strip of small widths. In this section, we study smaller values of $k$ on square lattice using Monte Carlo simulations. We present evidence,  from simulations at fixed density, of coexistence of nematic and HDD phases. 

\begin{figure}
\includegraphics[width=\columnwidth]{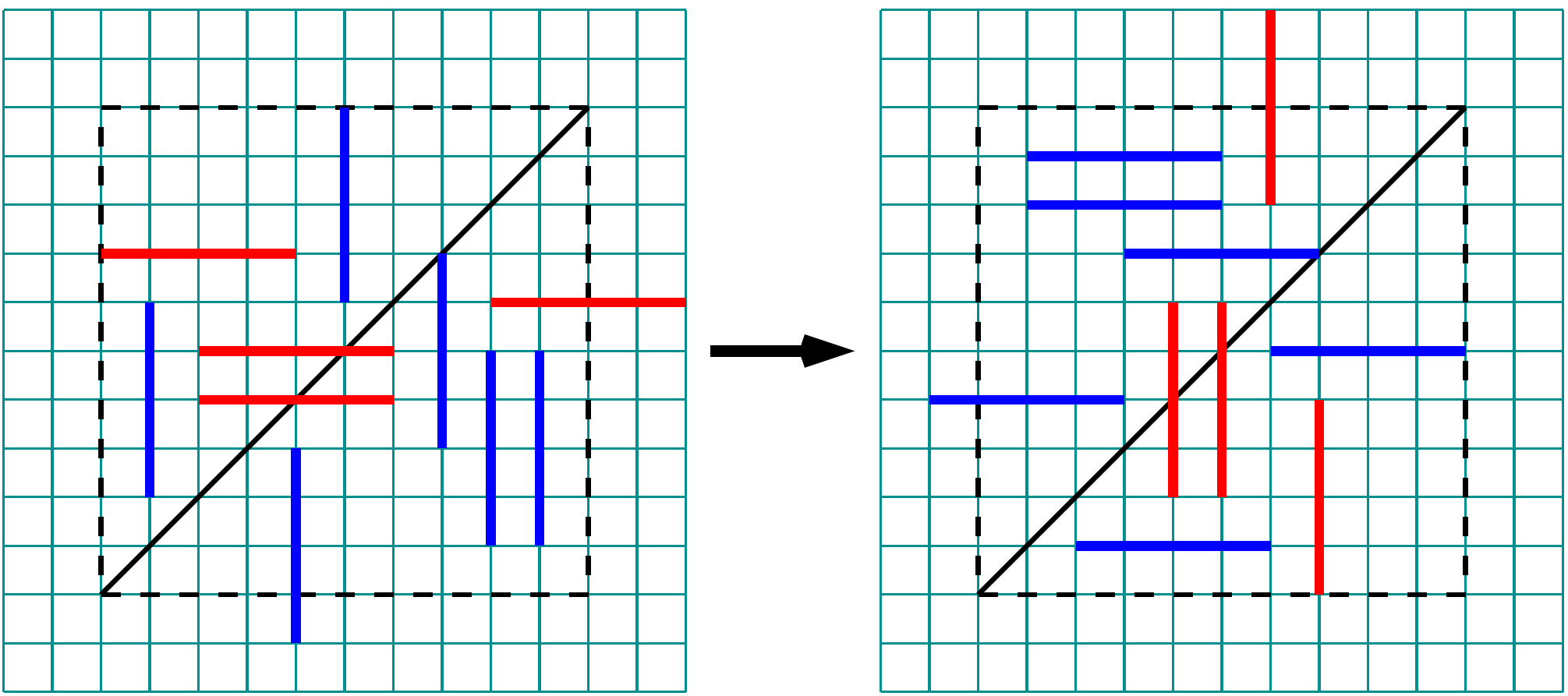}
\caption{\label{fig:rotation}  (Color online)  An example of the generalized flip used in the Monte Carlo simulations. A box of size $M\times M$, shown by dotted line, is chosen at random. All the rods are reflected about a randomly chosen diagonal (in this case $+\pi/4$).  }
\end{figure}

It is in general difficult to equilibrate the system of rods  at densities close to full packing. The algorithm that has been most efficient in equilibration is grand canonical in nature~\cite{2012-krds-aipcp-monte,2013-krds-pre-nematic}, but for fixed density simulations, we need an efficient algorithm that does not change the number of rods. To show phase coexistence, we need to work with  fixed number of rods.  For this, we were able to reach equilibration with the following  three moves:

{\it Generalized flip:} Choose a site at random and consider an $M\times M$ box whose left bottom corner is at the chosen site. Choose at random one of the diagonals from those in the $\pi/4$ or the $-\pi/4$ directions. Reflect all the rods whose center of mass lies within the box about the chosen diagonal. An example is shown in Fig.~\ref{fig:rotation}. If the reflected configuration does not violate the hard core constraint, it is accepted, else it is rejected. In our simulations, we choose $M=k$.

{\it Enhanced diffusion:} A row is chosen at random (a row could be in the horizontal or vertical direction). Suppose it is a horizontal row. Remove all $x$-mers such that the row breaks up into segments separated by $y$-mers. Each segment is re-populated with a new configuration with the same number of original $x$-mers in that segment. This new configuration is chosen at random from among all possible configurations.  The rearrangement of rods correspond to multiple diffusion moves in the direction of the rods' orientation. The procedure to generate a new configuration can be found in Refs.~\cite{jaleel2021rejection,jaleel2021hard}.

{\it Sliding:} The sliding move will correspond to the movement of entire rows of rods of one kind. Choose a row at random (say horizontal). If it is not fully nematic, then nothing is done. If nematic, choose a direction (up or down) at random and identify the contiguous nematic lines in this direction. These set of lines are shifted to the next available row in the same direction, which corresponds to a row where all the vertical links with bottom edge on the row are not occupied by rods. It can be checked that the move is reversible and satisfies detailed balance. The sliding move speeds up the aggregation of nematic lines. We find that in the absence of this move, the dynamics is very slow and the system does not equilibrate within our simulation times. 
\begin{figure}
\includegraphics[width=\columnwidth]{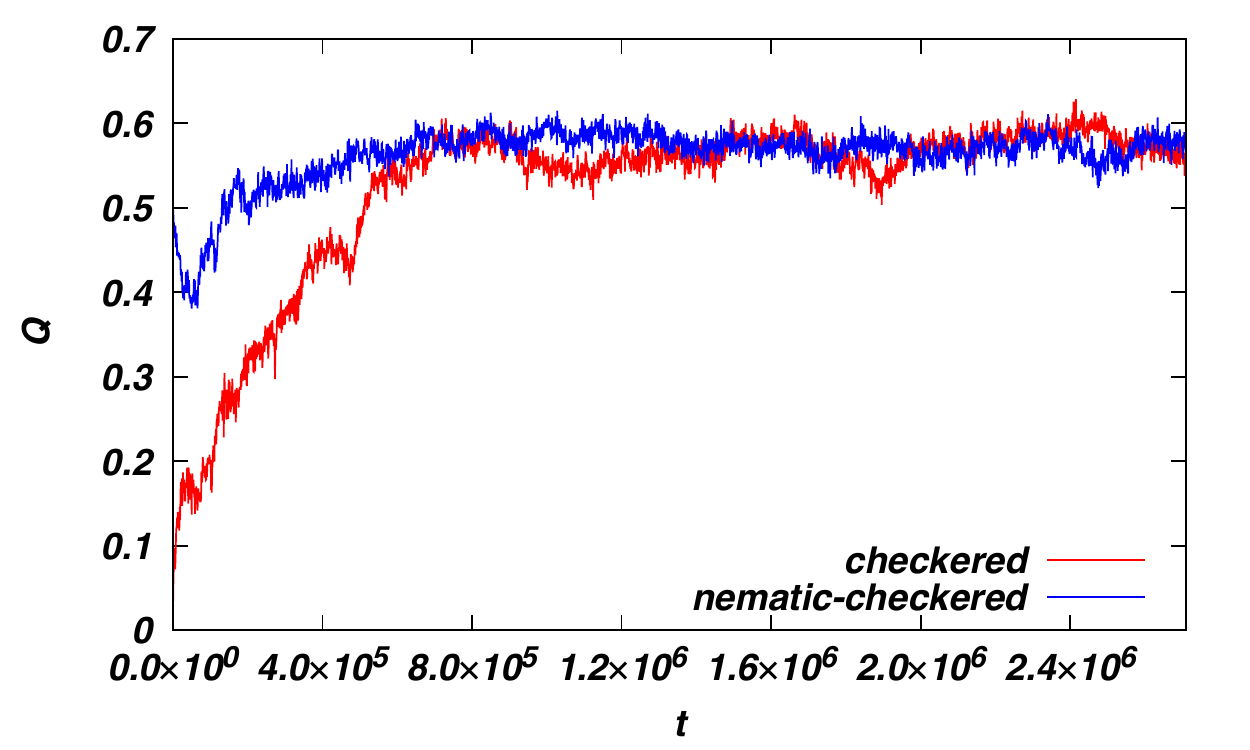}
\caption{\label{fig:mc_timeevol}  (Color online) Time evolution of the nematic order parameter $Q$ for two different initial conditions: checkered and nematic-checkered. The data are for $L=504$ and $\rho=0.968$. At long times, the nematic order is independent of initial condition. }
\end{figure}

We will define one Monte Carlo time step to correspond to $2 L$ row updates, $3 (L^2/M^2+1)$ flip moves and $2 L$ sliding moves. 

All the results that are presented are for $k=9$.
We first show that the Monte Carlo algorithm is able to equilibrate the system at the densities we are interested in. To do so, we consider two different initial conditions: (1) checkered, where the lattice is broken up into $k\times k$ plaquettes with a randomly chosen direction for rods in each plaquette, and (2) nematic-checkered, where half the rods are in a nematic phase and the other half in a checkered phase with the interface being perpendicular to the nematic order. Figure~\ref{fig:mc_timeevol} shows the time evolution of the nematic order parameter $Q(t)$ for a single realization for the two initial conditions for $L=504$ and $\rho=0.968$. After initial transients, $Q(t)$ fluctuates about a steady state  value that is independent of the initial conditions, showing that equilibration is achieved. 
\begin{figure*}
\includegraphics[width=0.8\textwidth]{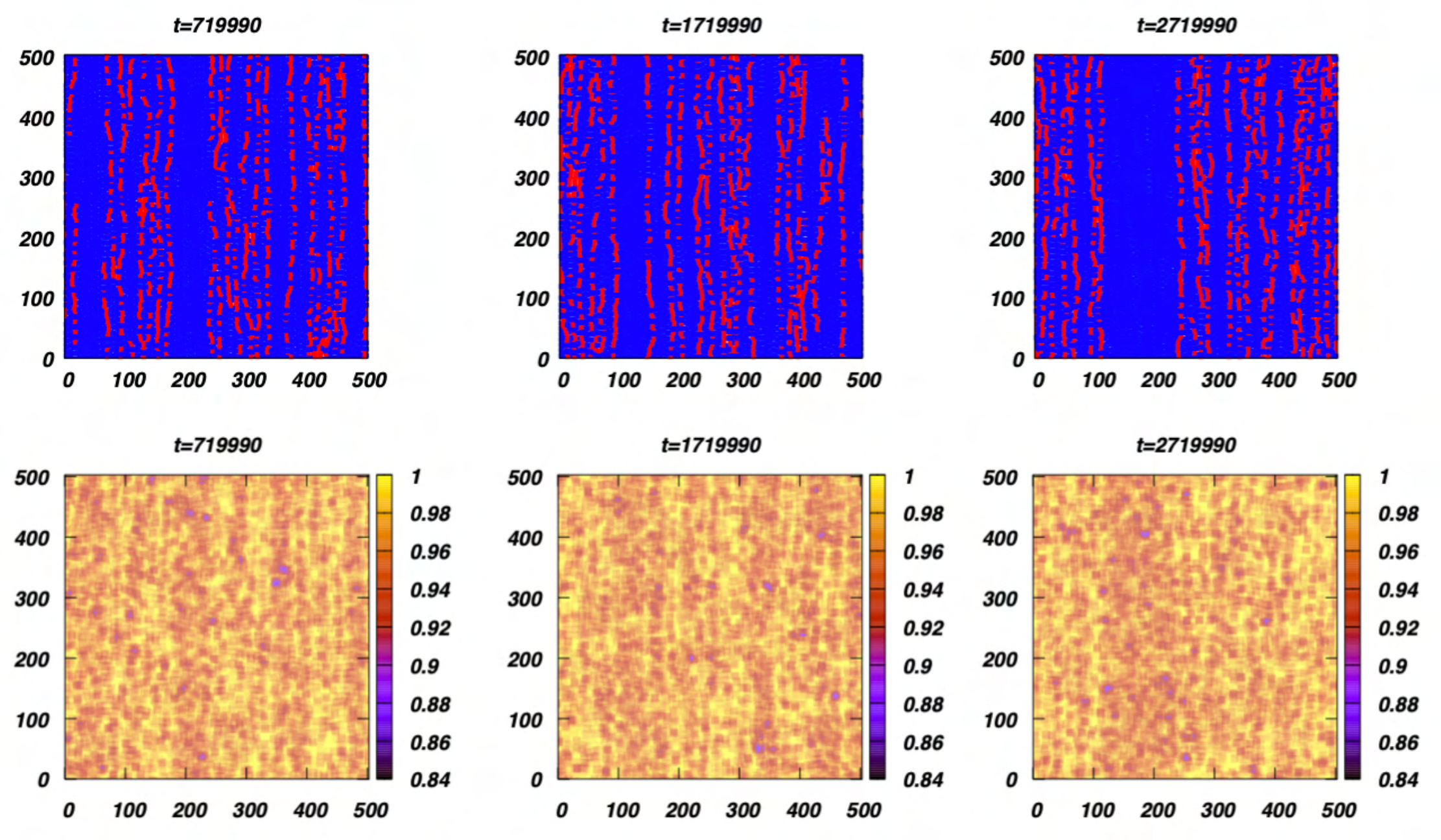}
\caption{\label{fig:snapshots}  (Color online) Snapshots of the configurations (top row) and the color map of the corresponding densities (bottom row) at different times. The initial phase was checkered. The blue rods are vertical and the red rods are horizontal. The data are for $L=504$, $\rho=0.968$ and $k=9$. At short times ($t=719990$), the system is homogeneous. At intermediate times ($t=1719990$), small nematic regions appears. At late times ($t=2719990$), the coexistence of a nematic phase with high density phase can be seen. The densities corresponding to the nematic phase show a lower value.  }
\end{figure*}

In Fig.~\ref{fig:snapshots}, we show snapshots of the configurations and the spatial variation of density for different times. The initial condition is checkered.  At initial times ($t=719990$ in Fig.~\ref{fig:snapshots}), the system is homogeneous. At intermediate times ($t=1719990$ in Fig.~\ref{fig:snapshots}), we see the formation of small nematic regions. At later times ($t=2719990$ in Fig.~\ref{fig:snapshots}), the nematic region becomes stable (also see below). The density in the nematic phase is lower than the other regions, as can be seen from the corresponding density maps. We thus conclude that the  system equilibrates in a phase-separated configuration characterised by co-existence of the nematic and high density phase, a strong signature of a first order transition.

\begin{figure}
\includegraphics[width=\columnwidth]{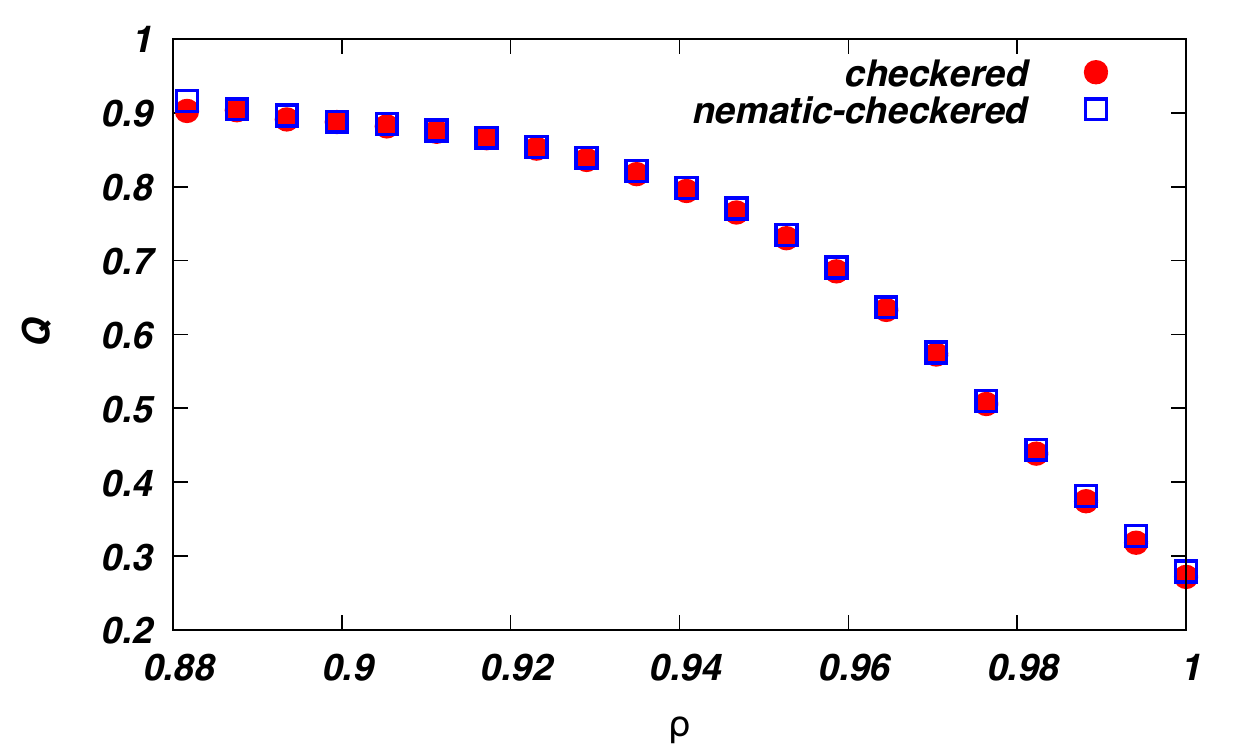}
\caption{\label{fig:mc_correlations}  (Color online) The local mean nematic order for a given local density $\rho$. The local $\rho$ and $Q$ are obtained by averaging over a box of size $13\times 13$. The data are for $L=504$ and $\rho=0.968$, and for two different initial conditions: one which is checkered and the other where a nematic phase and a checkered phase are separated by a straight interface. }
\end{figure}

We now show quantitatively that the low density regions in the snapshots in  Fig.~\ref{fig:snapshots} correspond to the nematic phase and the high density regions have no nematic order. To do so, we define coarse-grained densities and nematic order for each lattice site by averaging these quantities over a box of size $13\times 13$ centered about the lattice point.  Figure~\ref{fig:mc_correlations} shows the mean nematic order for a given local density  for the two initial conditions for $L=504$ and $\rho=0.968$. the data is averaged over the steady state  with an interval of $10^4$ Monte Carlo steps.  The data show that, in the coexistence regime, the local nematic order decreases sharply with increasing local density. 
\begin{figure*}
\includegraphics[width=0.95\textwidth]{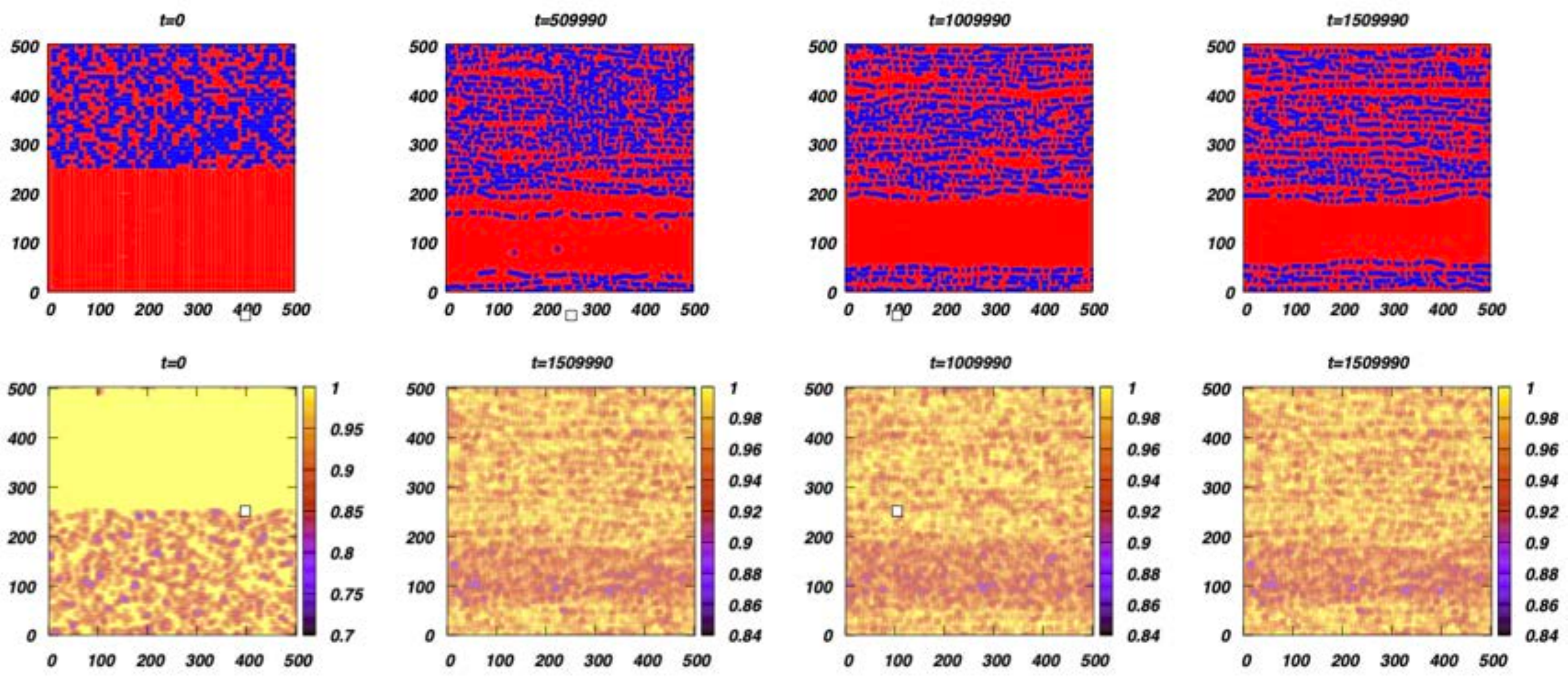}
\caption{\label{fig:snapshots1}  (Color online) Snapshots of the configurations (top row) and the color map of the corresponding densities (bottom row) at different times.  The blue rods are vertical and the red rods are horizontal. The initial configuration consists of a checkered phase of density one and a nematic phase as shown in the $t=0$ snapshot. The data are for $L=504$, $\rho=0.968$ and $k=9$ (same as Fig.~\ref{fig:snapshots}). With increasing time, the nematic region is stable. The densities corresponding to the nematic phase show a lower value. }
\end{figure*}

In the simulations, we find that the phase separation dynamics is very slow. To confirm that the phase-separated phase is stable, we do a reverse simulation for the same system as shown in Fig.~\ref{fig:snapshots}. The initial configuration is phase separated with one  half in a checkered phase with density one and the other half in a nematic phase with the interface parallel to the nematic orientation, as shown in $t=0$ snapshot in Fig.~\ref{fig:snapshots1}. 
We then evolve the system and check whether the phase separated phase persists. The snapshots at different times are shown in the top row of Fig.~\ref{fig:snapshots1}. As can be seen, the nematic region is stable. The corresponding density maps (bottom row of Fig.~\ref{fig:snapshots1}) show a lower density for the regions that are nematic. We conclude that, at this density, the system shows co-existence.

\section{\label{sec:conclusions} Summary and Discussion}

In this paper, we have argued that  in a system of monodispersed hard  rods on lattices, the  transition from the  nematic phase with orientational order to a  high density phase with no orientational order is a first order transition.  By estimating the entropy of the high density phase by counting over a subset of allowed configurations, we determine the large $k$ asymptotic behavior of the the critical chemical potential per rod at the transition, $\mu^*$,  and the jump in density at the transition, $\delta \epsilon$. We showed that  to leading order $\mu^* = k \ln [ k/\ln k]$ and $\delta \epsilon = \ln k/ k^2$.
These asymptotic behavior are shown to change by a very small amount [$\mathcal{O}(k^{-k})$] if the restrictions on the subset of configurations are  removed one by one. Thus, the asymptotic behavior is independent of the exact choice of the subset of configurations that we count.  We also obtained  the  solution of the problem on the $k \times \infty$ strip. The entropy on the strip at full packing is known to have the same asymptotic behavior of  entropy as  that for the square lattice~\cite{dhar2021entropy}. We showed that the entropy of the strip extrapolates smoothly from the value in the nematic phase to that in the high density phase consistent with our different estimates. While there is no transition on the strip, we see a sharp jump in the density at a value of the chemical potential.  This analysis is consistent with the hypothesis of  a first order transition in the two dimensional problem, which is smeared in  the strip.  Finally, we also presented evidence for the first order nature of the transition using Monte Carlo simulations for  $k =9$ on the square lattice.  Using a combination of enhanced diffusion, sliding moves when entire lines are displaced, and generalized flips when rods are rotated, we were able to equilibrate the systems at high densities. Phase separation was observed, which is a clear signature of a first order transition.

We note that when two different approximate equations are used
for the Gibbs free energy in the two different phases, 
their intersection gives  the location as well as the density jump in a  first order transition, independent of the details of the problem.  This would be the case, even if the actual transition is continuous. The special feature of the problem of hard rods considered in this paper, which makes our results more  trustworthy,  is the fact that the asymptotic behavior of the solution is unchanged on making improved approximations, and is robust against minor changes in the equations.

Within our approximation scheme, we find a first order transition, which implies that the Gibbs free energy per site is the same for a whole range  of density values.  If this degeneracy is lifted in the bulk free energy, even by a correction factor of order $k^{-k}$, the range of first order transition may be modified substantially, or the transition may disappear altogether.  This is what happens in the $k\times \infty$ strip, where there is no transition. However, if we assume that there is a transition in two dimensions, then the scenario presented in this paper is the simplest, and quite plausibly correct.  Making  these arguments  more rigorous  would be desirable.


The arguments based on different estimates for entropies do not depend on dimension. Hence, we expect that the transition from the nematic to the high density phase (layered disordered in three and higher dimensions~\cite{2017-vdr-jsm-different}) will be discontinuous in all dimensions. Demonstrating this in Monte Carlo simulations is a challenging problem.

Our arguments are easily extended to other lattices like the triangular lattice and the transition will be expected to be first order on these as well.  Earlier Monte Carlo simulations for $k=7$ on a triangular lattice indicated the transition to be continuous and consistent with the exponents of three state Potts model~\cite{2013-krds-pre-nematic}. That analysis (also for the square lattice) was based on the data then available, and perhaps did not reach true equilibrium due to slow phase separation kinetics.  We feel that the arguments presented in this paper are more convincing. Also, in Ref.~\cite{2013-krds-pre-nematic}, we had given some evidence of high density phase having power law correlations. It would be interesting to revisit the question of nature of correlations in the HDD phase with better simulations.  Preliminary data shows that the generalized flip implemented in this paper will decrease the autocorrelation time by a factor larger than 100, and thus the improved algorithm may be effective.
   
The arguments presented here for a first order transition will also apply to the phase transitions at high density in systems of hard rectangles of size  $\ell \times m $~\cite{2014-kr-pre-phase,2015-kr-epjb-phase,2015-kr-pre-asymptotic}. In these systems, there is a transition from a columnar phase, which has both orientational order as well as translational order in one direction, to a high density phase which has no nematic order and has sublattice order only if greatest common divisor of $\ell$ and $m$ is larger than one. The columnar entropy is well approximated by the one dimensional entropy. The description of the high density phase in this paper carries over to rectangles. For example, the entropy per site in the $m \times m k$ rectangle should be $1/m^2$ times the entropy per site in the $1 \times k$ rods~\cite{2014-kr-pre-phase}. In Ref.~\cite{2014-kr-pre-phase}, the transition for $2\times 10$ rectangles was found to be continuous in grand canonical Monte Carlo simulations. It would be of interest to re-examine this transition.

It may also be interesting to extend the study of the problem on  the $k \times \infty $ strip to bigger widths  using simulations.  It seems reasonable to expect that similar behavior to that seen for $k \times \infty$ strip would be seen in wider strips also, say of width $2k$.  One possible method to study strips is to use flat histogram techniques. Cluster algorithms on strips combined with Wang-Landau flat histogram algorithm have been recently successful in obtaining entropy of hard core lattice gas models even at full packing~\cite{jaleel2021rejection,jaleel2021hard}. Using $k\times L$ lattices to benchmark this algorithm, it would be an interesting direction for further study.


\begin{thebibliography}{33}%
\makeatletter
\providecommand \@ifxundefined [1]{%
 \@ifx{#1\undefined}
}%
\providecommand \@ifnum [1]{%
 \ifnum #1\expandafter \@firstoftwo
 \else \expandafter \@secondoftwo
 \fi
}%
\providecommand \@ifx [1]{%
 \ifx #1\expandafter \@firstoftwo
 \else \expandafter \@secondoftwo
 \fi
}%
\providecommand \natexlab [1]{#1}%
\providecommand \enquote  [1]{``#1''}%
\providecommand \bibnamefont  [1]{#1}%
\providecommand \bibfnamefont [1]{#1}%
\providecommand \citenamefont [1]{#1}%
\providecommand \href@noop [0]{\@secondoftwo}%
\providecommand \href [0]{\begingroup \@sanitize@url \@href}%
\providecommand \@href[1]{\@@startlink{#1}\@@href}%
\providecommand \@@href[1]{\endgroup#1\@@endlink}%
\providecommand \@sanitize@url [0]{\catcode `\\12\catcode `\$12\catcode
  `\&12\catcode `\#12\catcode `\^12\catcode `\_12\catcode `\%12\relax}%
\providecommand \@@startlink[1]{}%
\providecommand \@@endlink[0]{}%
\providecommand \url  [0]{\begingroup\@sanitize@url \@url }%
\providecommand \@url [1]{\endgroup\@href {#1}{\urlprefix }}%
\providecommand \urlprefix  [0]{URL }%
\providecommand \Eprint [0]{\href }%
\providecommand \doibase [0]{https://doi.org/}%
\providecommand \selectlanguage [0]{\@gobble}%
\providecommand \bibinfo  [0]{\@secondoftwo}%
\providecommand \bibfield  [0]{\@secondoftwo}%
\providecommand \translation [1]{[#1]}%
\providecommand \BibitemOpen [0]{}%
\providecommand \bibitemStop [0]{}%
\providecommand \bibitemNoStop [0]{.\EOS\space}%
\providecommand \EOS [0]{\spacefactor3000\relax}%
\providecommand \BibitemShut  [1]{\csname bibitem#1\endcsname}%
\let\auto@bib@innerbib\@empty
\bibitem [{\citenamefont {Onsager}(1949)}]{1949-o-nyas-effects}%
  \BibitemOpen
  \bibfield  {author} {\bibinfo {author} {\bibfnamefont {L.}~\bibnamefont
  {Onsager}},\ }\bibfield  {title} {\bibinfo {title} {The effects of shape on
  the interaction of colloidal particles},\ }\href
  {https://doi.org/10.1111/j.1749-6632.1949.tb27296.x} {\bibfield  {journal}
  {\bibinfo  {journal} {Ann. N. Y. Acad. Sci.}\ }\textbf {\bibinfo {volume}
  {51}},\ \bibinfo {pages} {627} (\bibinfo {year} {1949})}\BibitemShut
  {NoStop}%
\bibitem [{\citenamefont {Zwanzig}(1963)}]{1963-z-jcp-first}%
  \BibitemOpen
  \bibfield  {author} {\bibinfo {author} {\bibfnamefont {R.}~\bibnamefont
  {Zwanzig}},\ }\bibfield  {title} {\bibinfo {title} {First‐order phase
  transition in a gas of long thin rods},\ }\href@noop {} {\bibfield  {journal}
  {\bibinfo  {journal} {J. Chem. Phys.}\ }\textbf {\bibinfo {volume} {39}},\
  \bibinfo {pages} {1714} (\bibinfo {year} {1963})}\BibitemShut {NoStop}%
\bibitem [{\citenamefont {Fraden}\ \emph {et~al.}(1989)\citenamefont {Fraden},
  \citenamefont {Maret}, \citenamefont {Caspar},\ and\ \citenamefont
  {Meyer}}]{1989-fmcm-prl-isotropic}%
  \BibitemOpen
  \bibfield  {author} {\bibinfo {author} {\bibfnamefont {S.}~\bibnamefont
  {Fraden}}, \bibinfo {author} {\bibfnamefont {G.}~\bibnamefont {Maret}},
  \bibinfo {author} {\bibfnamefont {D.~L.~D.}\ \bibnamefont {Caspar}},\ and\
  \bibinfo {author} {\bibfnamefont {R.~B.}\ \bibnamefont {Meyer}},\ }\bibfield
  {title} {\bibinfo {title} {Isotropic-nematic phase transition and angular
  correlations in isotropic suspensions of tobacco mosaic virus},\ }\href
  {https://doi.org/10.1103/PhysRevLett.63.2068} {\bibfield  {journal} {\bibinfo
   {journal} {Phys. Rev. Lett.}\ }\textbf {\bibinfo {volume} {63}},\ \bibinfo
  {pages} {2068} (\bibinfo {year} {1989})}\BibitemShut {NoStop}%
\bibitem [{\citenamefont {de~Gennes}\ and\ \citenamefont
  {Prost}(1995)}]{1995-oup-gp-physics}%
  \BibitemOpen
  \bibfield  {author} {\bibinfo {author} {\bibfnamefont {P.~G.}\ \bibnamefont
  {de~Gennes}}\ and\ \bibinfo {author} {\bibfnamefont {J.}~\bibnamefont
  {Prost}},\ }\href@noop {} {\emph {\bibinfo {title} {The physics of liquid
  crystals}}},\ Vol.~\bibinfo {volume} {83}\ (\bibinfo  {publisher} {Oxford
  university press},\ \bibinfo {year} {1995})\BibitemShut {NoStop}%
\bibitem [{\citenamefont {Islam}\ \emph {et~al.}(2004)\citenamefont {Islam},
  \citenamefont {Alsayed}, \citenamefont {Dogic}, \citenamefont {Zhang},
  \citenamefont {Lubensky},\ and\ \citenamefont
  {Yodh}}]{2004-iadzly-prl-nematic}%
  \BibitemOpen
  \bibfield  {author} {\bibinfo {author} {\bibfnamefont {M.~F.}\ \bibnamefont
  {Islam}}, \bibinfo {author} {\bibfnamefont {A.~M.}\ \bibnamefont {Alsayed}},
  \bibinfo {author} {\bibfnamefont {Z.}~\bibnamefont {Dogic}}, \bibinfo
  {author} {\bibfnamefont {J.}~\bibnamefont {Zhang}}, \bibinfo {author}
  {\bibfnamefont {T.~C.}\ \bibnamefont {Lubensky}},\ and\ \bibinfo {author}
  {\bibfnamefont {A.~G.}\ \bibnamefont {Yodh}},\ }\bibfield  {title} {\bibinfo
  {title} {Nematic nanotube gels},\ }\href
  {https://doi.org/10.1103/PhysRevLett.92.088303} {\bibfield  {journal}
  {\bibinfo  {journal} {Phys. Rev. Lett.}\ }\textbf {\bibinfo {volume} {92}},\
  \bibinfo {pages} {088303} (\bibinfo {year} {2004})}\BibitemShut {NoStop}%
\bibitem [{\citenamefont {Ghosh}\ and\ \citenamefont
  {Dhar}(2007)}]{2007-gd-epl-on}%
  \BibitemOpen
  \bibfield  {author} {\bibinfo {author} {\bibfnamefont {A.}~\bibnamefont
  {Ghosh}}\ and\ \bibinfo {author} {\bibfnamefont {D.}~\bibnamefont {Dhar}},\
  }\bibfield  {title} {\bibinfo {title} {On the orientational ordering of long
  rods on a lattice},\ }\href {http://stacks.iop.org/0295-5075/78/i=2/a=20003}
  {\bibfield  {journal} {\bibinfo  {journal} {Europhys. Lett.}\ }\textbf
  {\bibinfo {volume} {78}},\ \bibinfo {pages} {20003} (\bibinfo {year}
  {2007})}\BibitemShut {NoStop}%
\bibitem [{\citenamefont {Matoz-Fernandez}\ \emph
  {et~al.}(2008{\natexlab{a}})\citenamefont {Matoz-Fernandez}, \citenamefont
  {Linares},\ and\ \citenamefont
  {Ramirez-Pastor}}]{2008-mlr-epl-determination}%
  \BibitemOpen
  \bibfield  {author} {\bibinfo {author} {\bibfnamefont {D.~A.}\ \bibnamefont
  {Matoz-Fernandez}}, \bibinfo {author} {\bibfnamefont {D.~H.}\ \bibnamefont
  {Linares}},\ and\ \bibinfo {author} {\bibfnamefont {A.~J.}\ \bibnamefont
  {Ramirez-Pastor}},\ }\bibfield  {title} {\bibinfo {title} {Determination of
  the critical exponents for the isotropic-nematic phase transition in a system
  of long rods on two-dimensional lattices: Universality of the transition},\
  }\href@noop {} {\bibfield  {journal} {\bibinfo  {journal} {Europhys. Lett.}\
  }\textbf {\bibinfo {volume} {82}},\ \bibinfo {pages} {50007} (\bibinfo {year}
  {2008}{\natexlab{a}})}\BibitemShut {NoStop}%
\bibitem [{\citenamefont {Matoz-Fernandez}\ \emph
  {et~al.}(2008{\natexlab{b}})\citenamefont {Matoz-Fernandez}, \citenamefont
  {Linares},\ and\ \citenamefont {Ramirez-Pastor}}]{2008-mlr-jcp-critical}%
  \BibitemOpen
  \bibfield  {author} {\bibinfo {author} {\bibfnamefont {D.~A.}\ \bibnamefont
  {Matoz-Fernandez}}, \bibinfo {author} {\bibfnamefont {D.~H.}\ \bibnamefont
  {Linares}},\ and\ \bibinfo {author} {\bibfnamefont {A.~J.}\ \bibnamefont
  {Ramirez-Pastor}},\ }\bibfield  {title} {\bibinfo {title} {Critical behavior
  of long straight rigid rods on two-dimensional lattices: Theory and monte
  carlo simulations},\ }\href {http://dx.doi.org/10.1063/1.2927877} {\bibfield
  {journal} {\bibinfo  {journal} {J. Chem. Phys.}\ }\textbf {\bibinfo {volume}
  {128}},\ \bibinfo {pages} {214902} (\bibinfo {year}
  {2008}{\natexlab{b}})}\BibitemShut {NoStop}%
\bibitem [{\citenamefont {Fischer}\ and\ \citenamefont
  {Vink}(2009)}]{2009-fv-epl-restricted}%
  \BibitemOpen
  \bibfield  {author} {\bibinfo {author} {\bibfnamefont {T.}~\bibnamefont
  {Fischer}}\ and\ \bibinfo {author} {\bibfnamefont {R.~L.~C.}\ \bibnamefont
  {Vink}},\ }\bibfield  {title} {\bibinfo {title} {Restricted orientation
  "liquid crystal" in two dimensions: Isotropic-nematic transition or
  liquid-gas one(?)},\ }\href {http://stacks.iop.org/0295-5075/85/i=5/a=56003}
  {\bibfield  {journal} {\bibinfo  {journal} {Europhys. Lett.}\ }\textbf
  {\bibinfo {volume} {85}},\ \bibinfo {pages} {56003} (\bibinfo {year}
  {2009})}\BibitemShut {NoStop}%
\bibitem [{\citenamefont {Matoz-Fernandez}\ \emph
  {et~al.}(2008{\natexlab{c}})\citenamefont {Matoz-Fernandez}, \citenamefont
  {Linares},\ and\ \citenamefont {Ramirez-Pastor}}]{2008-mlr-pa-critical}%
  \BibitemOpen
  \bibfield  {author} {\bibinfo {author} {\bibfnamefont {D.}~\bibnamefont
  {Matoz-Fernandez}}, \bibinfo {author} {\bibfnamefont {D.}~\bibnamefont
  {Linares}},\ and\ \bibinfo {author} {\bibfnamefont {A.}~\bibnamefont
  {Ramirez-Pastor}},\ }\bibfield  {title} {\bibinfo {title} {Critical behavior
  of long linear k-mers on honeycomb lattices},\ }\href
  {http://www.sciencedirect.com/science/article/pii/S0378437108007127}
  {\bibfield  {journal} {\bibinfo  {journal} {Physica A}\ }\textbf {\bibinfo
  {volume} {387}},\ \bibinfo {pages} {6513 } (\bibinfo {year}
  {2008}{\natexlab{c}})}\BibitemShut {NoStop}%
\bibitem [{\citenamefont {Disertori}\ and\ \citenamefont
  {Giuliani}(2013)}]{2013-dg-cmp-nematic}%
  \BibitemOpen
  \bibfield  {author} {\bibinfo {author} {\bibfnamefont {M.}~\bibnamefont
  {Disertori}}\ and\ \bibinfo {author} {\bibfnamefont {A.}~\bibnamefont
  {Giuliani}},\ }\bibfield  {title} {\bibinfo {title} {The nematic phase of a
  system of long hard rods},\ }\href
  {https://doi.org/10.1007/s00220-013-1767-1} {\bibfield  {journal} {\bibinfo
  {journal} {Commun. Math. Phys.}\ }\textbf {\bibinfo {volume} {323}},\
  \bibinfo {pages} {143} (\bibinfo {year} {2013})}\BibitemShut {NoStop}%
\bibitem [{\citenamefont {Dhar}\ \emph {et~al.}(2011)\citenamefont {Dhar},
  \citenamefont {Rajesh},\ and\ \citenamefont {Stilck}}]{2011-drs-pre-hard}%
  \BibitemOpen
  \bibfield  {author} {\bibinfo {author} {\bibfnamefont {D.}~\bibnamefont
  {Dhar}}, \bibinfo {author} {\bibfnamefont {R.}~\bibnamefont {Rajesh}},\ and\
  \bibinfo {author} {\bibfnamefont {J.~F.}\ \bibnamefont {Stilck}},\ }\bibfield
   {title} {\bibinfo {title} {Hard rigid rods on a bethe-like lattice},\ }\href
  {https://doi.org/10.1103/PhysRevE.84.011140} {\bibfield  {journal} {\bibinfo
  {journal} {Phys. Rev. E}\ }\textbf {\bibinfo {volume} {84}},\ \bibinfo
  {pages} {011140} (\bibinfo {year} {2011})}\BibitemShut {NoStop}%
\bibitem [{\citenamefont {Padavala}\ \emph {et~al.}(2021)\citenamefont
  {Padavala}, \citenamefont {Singh},\ and\ \citenamefont
  {Kundu}}]{padavala2021machine}%
  \BibitemOpen
  \bibfield  {author} {\bibinfo {author} {\bibfnamefont {K.}~\bibnamefont
  {Padavala}}, \bibinfo {author} {\bibfnamefont {A.}~\bibnamefont {Singh}},\
  and\ \bibinfo {author} {\bibfnamefont {J.}~\bibnamefont {Kundu}},\ }\bibfield
   {title} {\bibinfo {title} {Machine learned phase transitions in a system of
  anisotropic particles on a square lattice},\ }\href@noop {} {\bibfield
  {journal} {\bibinfo  {journal} {arXiv preprint arXiv:2102.03006}\ } (\bibinfo
  {year} {2021})}\BibitemShut {NoStop}%
\bibitem [{\citenamefont {Kundu}\ \emph {et~al.}(2012)\citenamefont {Kundu},
  \citenamefont {Rajesh}, \citenamefont {Dhar},\ and\ \citenamefont
  {Stilck}}]{2012-krds-aipcp-monte}%
  \BibitemOpen
  \bibfield  {author} {\bibinfo {author} {\bibfnamefont {J.}~\bibnamefont
  {Kundu}}, \bibinfo {author} {\bibfnamefont {R.}~\bibnamefont {Rajesh}},
  \bibinfo {author} {\bibfnamefont {D.}~\bibnamefont {Dhar}},\ and\ \bibinfo
  {author} {\bibfnamefont {J.~F.}\ \bibnamefont {Stilck}},\ }\bibfield  {title}
  {\bibinfo {title} {A monte carlo algorithm for studying phase transition in
  systems of hard rigid rods},\ }\href {https://doi.org/10.1063/1.4709907}
  {\bibfield  {journal} {\bibinfo  {journal} {AIP Conf. Proc.}\ }\textbf
  {\bibinfo {volume} {1447}},\ \bibinfo {pages} {113} (\bibinfo {year}
  {2012})}\BibitemShut {NoStop}%
\bibitem [{\citenamefont {Kundu}\ \emph {et~al.}(2013)\citenamefont {Kundu},
  \citenamefont {Rajesh}, \citenamefont {Dhar},\ and\ \citenamefont
  {Stilck}}]{2013-krds-pre-nematic}%
  \BibitemOpen
  \bibfield  {author} {\bibinfo {author} {\bibfnamefont {J.}~\bibnamefont
  {Kundu}}, \bibinfo {author} {\bibfnamefont {R.}~\bibnamefont {Rajesh}},
  \bibinfo {author} {\bibfnamefont {D.}~\bibnamefont {Dhar}},\ and\ \bibinfo
  {author} {\bibfnamefont {J.~F.}\ \bibnamefont {Stilck}},\ }\bibfield  {title}
  {\bibinfo {title} {Nematic-disordered phase transition in systems of long
  rigid rods on two-dimensional lattices},\ }\href
  {https://doi.org/10.1103/PhysRevE.87.032103} {\bibfield  {journal} {\bibinfo
  {journal} {Phys. Rev. E}\ }\textbf {\bibinfo {volume} {87}},\ \bibinfo
  {pages} {032103} (\bibinfo {year} {2013})}\BibitemShut {NoStop}%
\bibitem [{\citenamefont {Kundu}\ and\ \citenamefont
  {Rajesh}(2013)}]{2013-kr-pre-reentrant}%
  \BibitemOpen
  \bibfield  {author} {\bibinfo {author} {\bibfnamefont {J.}~\bibnamefont
  {Kundu}}\ and\ \bibinfo {author} {\bibfnamefont {R.}~\bibnamefont {Rajesh}},\
  }\bibfield  {title} {\bibinfo {title} {Reentrant disordered phase in a system
  of repulsive rods on a bethe-like lattice},\ }\href
  {https://doi.org/10.1103/PhysRevE.88.012134} {\bibfield  {journal} {\bibinfo
  {journal} {Phys. Rev. E}\ }\textbf {\bibinfo {volume} {88}},\ \bibinfo
  {pages} {012134} (\bibinfo {year} {2013})}\BibitemShut {NoStop}%
\bibitem [{\citenamefont {Vogel}\ \emph {et~al.}(2017)\citenamefont {Vogel},
  \citenamefont {Saravia},\ and\ \citenamefont
  {Ramirez-Pastor}}]{vogel2017phase}%
  \BibitemOpen
  \bibfield  {author} {\bibinfo {author} {\bibfnamefont {E.~E.}\ \bibnamefont
  {Vogel}}, \bibinfo {author} {\bibfnamefont {G.}~\bibnamefont {Saravia}},\
  and\ \bibinfo {author} {\bibfnamefont {A.~J.}\ \bibnamefont
  {Ramirez-Pastor}},\ }\bibfield  {title} {\bibinfo {title} {Phase transitions
  in a system of long rods on two-dimensional lattices by means of information
  theory},\ }\href@noop {} {\bibfield  {journal} {\bibinfo  {journal} {Phys.
  Rev. E}\ }\textbf {\bibinfo {volume} {96}},\ \bibinfo {pages} {062133}
  (\bibinfo {year} {2017})}\BibitemShut {NoStop}%
\bibitem [{\citenamefont {Vogel}\ \emph {et~al.}(2020)\citenamefont {Vogel},
  \citenamefont {Saravia}, \citenamefont {Ramirez-Pastor},\ and\ \citenamefont
  {Pasinetti}}]{vogel2020alternative}%
  \BibitemOpen
  \bibfield  {author} {\bibinfo {author} {\bibfnamefont {E.~E.}\ \bibnamefont
  {Vogel}}, \bibinfo {author} {\bibfnamefont {G.}~\bibnamefont {Saravia}},
  \bibinfo {author} {\bibfnamefont {A.~J.}\ \bibnamefont {Ramirez-Pastor}},\
  and\ \bibinfo {author} {\bibfnamefont {M.}~\bibnamefont {Pasinetti}},\
  }\bibfield  {title} {\bibinfo {title} {Alternative characterization of the
  nematic transition in deposition of rods on two-dimensional lattices},\
  }\href@noop {} {\bibfield  {journal} {\bibinfo  {journal} {Phys. Rev. E}\
  }\textbf {\bibinfo {volume} {101}},\ \bibinfo {pages} {022104} (\bibinfo
  {year} {2020})}\BibitemShut {NoStop}%
\bibitem [{\citenamefont {Chatelain}\ and\ \citenamefont
  {Gendiar}(2020)}]{chatelain2020absence}%
  \BibitemOpen
  \bibfield  {author} {\bibinfo {author} {\bibfnamefont {C.}~\bibnamefont
  {Chatelain}}\ and\ \bibinfo {author} {\bibfnamefont {A.}~\bibnamefont
  {Gendiar}},\ }\bibfield  {title} {\bibinfo {title} {Absence of logarithmic
  divergence of the entanglement entropies at the phase transitions of a 2d
  classical hard rod model},\ }\href@noop {} {\bibfield  {journal} {\bibinfo
  {journal} {Eur. Phys. J. B}\ }\textbf {\bibinfo {volume} {93}},\ \bibinfo
  {pages} {134} (\bibinfo {year} {2020})}\BibitemShut {NoStop}%
\bibitem [{\citenamefont {Vigneshwar}\ \emph {et~al.}(2017)\citenamefont
  {Vigneshwar}, \citenamefont {Dhar},\ and\ \citenamefont
  {Rajesh}}]{2017-vdr-jsm-different}%
  \BibitemOpen
  \bibfield  {author} {\bibinfo {author} {\bibfnamefont {N.}~\bibnamefont
  {Vigneshwar}}, \bibinfo {author} {\bibfnamefont {D.}~\bibnamefont {Dhar}},\
  and\ \bibinfo {author} {\bibfnamefont {R.}~\bibnamefont {Rajesh}},\
  }\bibfield  {title} {\bibinfo {title} {Different phases of a system of hard
  rods on three dimensional cubic lattice},\ }\href
  {http://stacks.iop.org/1742-5468/2017/i=11/a=113304} {\bibfield  {journal}
  {\bibinfo  {journal} {J. Stat. Mech.}\ }\textbf {\bibinfo {volume} {2017}},\
  \bibinfo {pages} {113304} (\bibinfo {year} {2017})}\BibitemShut {NoStop}%
\bibitem [{\citenamefont {Gschwind}\ \emph {et~al.}(2017)\citenamefont
  {Gschwind}, \citenamefont {Klopotek}, \citenamefont {Ai},\ and\ \citenamefont
  {Oettel}}]{2017-gkao-pre-isotropic}%
  \BibitemOpen
  \bibfield  {author} {\bibinfo {author} {\bibfnamefont {A.}~\bibnamefont
  {Gschwind}}, \bibinfo {author} {\bibfnamefont {M.}~\bibnamefont {Klopotek}},
  \bibinfo {author} {\bibfnamefont {Y.}~\bibnamefont {Ai}},\ and\ \bibinfo
  {author} {\bibfnamefont {M.}~\bibnamefont {Oettel}},\ }\bibfield  {title}
  {\bibinfo {title} {Isotropic-nematic transition for hard rods on a
  three-dimensional cubic lattice},\ }\href
  {https://doi.org/10.1103/PhysRevE.96.012104} {\bibfield  {journal} {\bibinfo
  {journal} {Phys. Rev. E}\ }\textbf {\bibinfo {volume} {96}},\ \bibinfo
  {pages} {012104} (\bibinfo {year} {2017})}\BibitemShut {NoStop}%
\bibitem [{\citenamefont {Kasteleyn}(1961)}]{1961-k-physica-statistics}%
  \BibitemOpen
  \bibfield  {author} {\bibinfo {author} {\bibfnamefont {P.}~\bibnamefont
  {Kasteleyn}},\ }\bibfield  {title} {\bibinfo {title} {The statistics of
  dimers on a lattice},\ }\href
  {https://doi.org/http://dx.doi.org/10.1016/0031-8914(61)90063-5} {\bibfield
  {journal} {\bibinfo  {journal} {Physica}\ }\textbf {\bibinfo {volume} {27}},\
  \bibinfo {pages} {1209 } (\bibinfo {year} {1961})}\BibitemShut {NoStop}%
\bibitem [{\citenamefont {Kasteleyn}(1963)}]{kasteleyn1963dimer}%
  \BibitemOpen
  \bibfield  {author} {\bibinfo {author} {\bibfnamefont {P.~W.}\ \bibnamefont
  {Kasteleyn}},\ }\bibfield  {title} {\bibinfo {title} {Dimer statistics and
  phase transitions},\ }\href@noop {} {\bibfield  {journal} {\bibinfo
  {journal} {J. Math. Phys.}\ }\textbf {\bibinfo {volume} {4}},\ \bibinfo
  {pages} {287} (\bibinfo {year} {1963})}\BibitemShut {NoStop}%
\bibitem [{\citenamefont {Fisher}\ and\ \citenamefont
  {Stephenson}(1963)}]{1963-fs-pr-statistical}%
  \BibitemOpen
  \bibfield  {author} {\bibinfo {author} {\bibfnamefont {M.~E.}\ \bibnamefont
  {Fisher}}\ and\ \bibinfo {author} {\bibfnamefont {J.}~\bibnamefont
  {Stephenson}},\ }\bibfield  {title} {\bibinfo {title} {Statistical mechanics
  of dimers on a plane lattice. ii. dimer correlations and monomers},\ }\href
  {https://doi.org/10.1103/PhysRev.132.1411} {\bibfield  {journal} {\bibinfo
  {journal} {Phys. Rev.}\ }\textbf {\bibinfo {volume} {132}},\ \bibinfo {pages}
  {1411} (\bibinfo {year} {1963})}\BibitemShut {NoStop}%
\bibitem [{\citenamefont {Ghosh}\ \emph {et~al.}(2007)\citenamefont {Ghosh},
  \citenamefont {Dhar},\ and\ \citenamefont {Jacobsen}}]{2007-gdj-pre-random}%
  \BibitemOpen
  \bibfield  {author} {\bibinfo {author} {\bibfnamefont {A.}~\bibnamefont
  {Ghosh}}, \bibinfo {author} {\bibfnamefont {D.}~\bibnamefont {Dhar}},\ and\
  \bibinfo {author} {\bibfnamefont {J.~L.}\ \bibnamefont {Jacobsen}},\
  }\bibfield  {title} {\bibinfo {title} {Random trimer tilings},\ }\href
  {https://doi.org/10.1103/PhysRevE.75.011115} {\bibfield  {journal} {\bibinfo
  {journal} {Phys. Rev. E}\ }\textbf {\bibinfo {volume} {75}},\ \bibinfo
  {pages} {011115} (\bibinfo {year} {2007})}\BibitemShut {NoStop}%
\bibitem [{\citenamefont {Kenyon}(2000)}]{kenyon2000conformal}%
  \BibitemOpen
  \bibfield  {author} {\bibinfo {author} {\bibfnamefont {R.}~\bibnamefont
  {Kenyon}},\ }\bibfield  {title} {\bibinfo {title} {Conformal invariance of
  domino tiling},\ }\href@noop {} {\bibfield  {journal} {\bibinfo  {journal}
  {Ann. Prob.}\ }\textbf {\bibinfo {volume} {28}},\ \bibinfo {pages} {759}
  (\bibinfo {year} {2000})}\BibitemShut {NoStop}%
\bibitem [{\citenamefont {Huse}\ \emph {et~al.}(2003)\citenamefont {Huse},
  \citenamefont {Krauth}, \citenamefont {Moessner},\ and\ \citenamefont
  {Sondhi}}]{2003-hkms-prl-coulomb}%
  \BibitemOpen
  \bibfield  {author} {\bibinfo {author} {\bibfnamefont {D.~A.}\ \bibnamefont
  {Huse}}, \bibinfo {author} {\bibfnamefont {W.}~\bibnamefont {Krauth}},
  \bibinfo {author} {\bibfnamefont {R.}~\bibnamefont {Moessner}},\ and\
  \bibinfo {author} {\bibfnamefont {S.~L.}\ \bibnamefont {Sondhi}},\ }\bibfield
   {title} {\bibinfo {title} {Coulomb and liquid dimer models in three
  dimensions},\ }\href {https://doi.org/10.1103/PhysRevLett.91.167004}
  {\bibfield  {journal} {\bibinfo  {journal} {Phys. Rev. Lett.}\ }\textbf
  {\bibinfo {volume} {91}},\ \bibinfo {pages} {167004} (\bibinfo {year}
  {2003})}\BibitemShut {NoStop}%
\bibitem [{\citenamefont {Dhar}\ and\ \citenamefont
  {Rajesh}(2021)}]{dhar2021entropy}%
  \BibitemOpen
  \bibfield  {author} {\bibinfo {author} {\bibfnamefont {D.}~\bibnamefont
  {Dhar}}\ and\ \bibinfo {author} {\bibfnamefont {R.}~\bibnamefont {Rajesh}},\
  }\bibfield  {title} {\bibinfo {title} {Entropy of fully packed hard rigid
  rods on d-dimensional hypercubic lattices},\ }\href@noop {} {\bibfield
  {journal} {\bibinfo  {journal} {Phys. Rev. E}\ }\textbf {\bibinfo {volume}
  {103}},\ \bibinfo {pages} {042130} (\bibinfo {year} {2021})}\BibitemShut
  {NoStop}%
\bibitem [{\citenamefont {Jaleel}\ \emph
  {et~al.}(2021{\natexlab{a}})\citenamefont {Jaleel}, \citenamefont {Thomas},
  \citenamefont {Mandal}, \citenamefont {Sumedha},\ and\ \citenamefont
  {Rajesh}}]{jaleel2021rejection}%
  \BibitemOpen
  \bibfield  {author} {\bibinfo {author} {\bibfnamefont {A.~A.~A.}\
  \bibnamefont {Jaleel}}, \bibinfo {author} {\bibfnamefont {J.~E.}\
  \bibnamefont {Thomas}}, \bibinfo {author} {\bibfnamefont {D.}~\bibnamefont
  {Mandal}}, \bibinfo {author} {\bibnamefont {Sumedha}},\ and\ \bibinfo
  {author} {\bibfnamefont {R.}~\bibnamefont {Rajesh}},\ }\bibfield  {title}
  {\bibinfo {title} {Rejection free cluster wang landau algorithm for hard core
  lattice gases},\ }\href@noop {} {\bibfield  {journal} {\bibinfo  {journal}
  {arXiv preprint arXiv:2108.01402}\ } (\bibinfo {year}
  {2021}{\natexlab{a}})}\BibitemShut {NoStop}%
\bibitem [{\citenamefont {Jaleel}\ \emph
  {et~al.}(2021{\natexlab{b}})\citenamefont {Jaleel}, \citenamefont {Mandal},\
  and\ \citenamefont {Rajesh}}]{jaleel2021hard}%
  \BibitemOpen
  \bibfield  {author} {\bibinfo {author} {\bibfnamefont {A.~A.~A.}\
  \bibnamefont {Jaleel}}, \bibinfo {author} {\bibfnamefont {D.}~\bibnamefont
  {Mandal}},\ and\ \bibinfo {author} {\bibfnamefont {R.}~\bibnamefont
  {Rajesh}},\ }\bibfield  {title} {\bibinfo {title} {Hard core lattice gas with
  third next-nearest neighbor exclusion on triangular lattice: one or two phase
  transitions?},\ }\href@noop {} {\bibfield  {journal} {\bibinfo  {journal}
  {arXiv preprint arXiv:2108.03547}\ } (\bibinfo {year}
  {2021}{\natexlab{b}})}\BibitemShut {NoStop}%
\bibitem [{\citenamefont {Kundu}\ and\ \citenamefont
  {Rajesh}(2014)}]{2014-kr-pre-phase}%
  \BibitemOpen
  \bibfield  {author} {\bibinfo {author} {\bibfnamefont {J.}~\bibnamefont
  {Kundu}}\ and\ \bibinfo {author} {\bibfnamefont {R.}~\bibnamefont {Rajesh}},\
  }\bibfield  {title} {\bibinfo {title} {Phase transitions in a system of hard
  rectangles on the square lattice},\ }\href
  {https://doi.org/10.1103/PhysRevE.89.052124} {\bibfield  {journal} {\bibinfo
  {journal} {Phys. Rev. E}\ }\textbf {\bibinfo {volume} {89}},\ \bibinfo
  {pages} {052124} (\bibinfo {year} {2014})}\BibitemShut {NoStop}%
\bibitem [{\citenamefont {Kundu}\ and\ \citenamefont
  {Rajesh}(2015{\natexlab{a}})}]{2015-kr-epjb-phase}%
  \BibitemOpen
  \bibfield  {author} {\bibinfo {author} {\bibfnamefont {J.}~\bibnamefont
  {Kundu}}\ and\ \bibinfo {author} {\bibfnamefont {R.}~\bibnamefont {Rajesh}},\
  }\bibfield  {title} {\bibinfo {title} {Phase transitions in systems of hard
  rectangles with non-integer aspect ratio},\ }\href
  {https://doi.org/10.1140/epjb/e2015-60210-7} {\bibfield  {journal} {\bibinfo
  {journal} {Eur. Phys. J. B}\ }\textbf {\bibinfo {volume} {88}},\ \bibinfo
  {pages} {133} (\bibinfo {year} {2015}{\natexlab{a}})}\BibitemShut {NoStop}%
\bibitem [{\citenamefont {Kundu}\ and\ \citenamefont
  {Rajesh}(2015{\natexlab{b}})}]{2015-kr-pre-asymptotic}%
  \BibitemOpen
  \bibfield  {author} {\bibinfo {author} {\bibfnamefont {J.}~\bibnamefont
  {Kundu}}\ and\ \bibinfo {author} {\bibfnamefont {R.}~\bibnamefont {Rajesh}},\
  }\bibfield  {title} {\bibinfo {title} {Asymptotic behavior of the
  isotropic-nematic and nematic-columnar phase boundaries for the system of
  hard rectangles on a square lattice},\ }\href
  {https://doi.org/10.1103/PhysRevE.91.012105} {\bibfield  {journal} {\bibinfo
  {journal} {Phys. Rev. E}\ }\textbf {\bibinfo {volume} {91}},\ \bibinfo
  {pages} {012105} (\bibinfo {year} {2015}{\natexlab{b}})}\BibitemShut
  {NoStop}%
\end{thebibliography}
%

\end{document}